\documentclass[twocolumn,letter]{IEEEtran}
\usepackage[usenames,dvipsnames]{color}
\usepackage{subfigure}
\usepackage{graphicx}
\usepackage{amsmath}
\usepackage{color}
\usepackage{multicol}
\usepackage{amssymb}

\usepackage{balance}
\usepackage{amsfonts}%
\usepackage{verbatim}%
\usepackage{enumerate}%
\usepackage{psfrag}
\usepackage{epsfig}
\usepackage{multicol}
\usepackage{setspace}
\usepackage{dsfont}

\usepackage{algorithm}
\usepackage{algorithmic}
\usepackage{cite}
\usepackage{float}
\usepackage{cite}
 
\begin{document}
\title{Maximum Secondary Stable Throughput of a Cooperative Secondary Transmitter-Receiver Pair: Protocol Design and Stability Analysis}
\author{ Ahmed El Shafie$^\dagger$, Tamer Khattab$^*$, Amr El-Keyi$^\dagger$, Mohamed Nafie$^\dagger$\\
\small \begin{tabular}{c}
$^\dagger$Wireless Intelligent Networks Center (WINC), Nile University, Giza, Egypt. \\
$^*$Electrical Engineering, Qatar University, Doha, Qatar. \\
\end{tabular}
\thanks{Part of this work has been presented in the 8th International Conference on Cognitive Radio Oriented Wireless Networks (CROWNCOM), 2013 \cite{6636823}}.}

\date{}
\maketitle
\thispagestyle{empty}
\pagestyle{empty}
\begin{abstract}
In this paper, we investigate the impact of cooperation between a secondary transmitter-receiver pair and a primary transmitter (PT) on the maximum stable throughput of the primary-secondary network. Each transmitter, primary or secondary, has a buffer for storing its own traffic. In addition to its own buffer, the secondary transmitter (ST) has a buffer for storing a fraction of the undelivered primary packets due to channel impairments. Moreover, the secondary destination has a relaying queue for storing a fraction of the undelivered primary packets. In the proposed cooperative system, the ST and the secondary destination increase the spectrum availability for the secondary packets by relaying the unsuccessfully transmitted packets of the PT. We consider two multiple access strategies to be used by the ST and the secondary destination to utilize the silence sessions of the PT. Numerical results demonstrate the gains of the proposed cooperative system over the non-cooperation case.
\end{abstract}
\begin{IEEEkeywords}
Cognitive radio, queues, stability region, inner and outer bounds, dominant system.
\end{IEEEkeywords}


\maketitle
\thispagestyle{empty}
\pagestyle{empty}
\section{Introduction}

\IEEEPARstart{T}\small{h}e electromagnetic radio spectrum is a precious resource, whose use is licensed by governments \cite{haykin2005cognitive}. Regulatory bodies have come to realize that most of the time, large portions of certain licensed frequency bands remain
unused. One of the most efficient ways to increase the spectrum usage is to use a secondary system that overlaps with the primary licensed system. The intuitive intention behind secondary spectrum licensing is to efficiently increase the spectral
usage of the network while, depending on the type of licensing, not perturbing
the higher priority users (primary users). Cognitive radio (CR) systems are seen as a candidate prime solution that can significantly mitigate the current low spectral efficiency in the electromagnetic spectrum. A CR system is defined as an {\it intelligent} wireless communication system that is fully aware of its environment and uses methodologies of learning and reasoning in order to
dynamically adapt its transmission parameters, e.g., operating spectrum, modulation schemes, coding, and transmission power, to access portions of spectrum by exploiting the existence of spectrum holes left unused by a primary system.

Cooperative diversity is a recently emerged technique
for wireless communications that has gained wide attention
\cite{sadek}. Most of the work on cooperative
communications has concentrated on the physical layer (PHY) aspects
of the problem by improving transmission parameters at the PHY. In a wireless communication network with many source-receiver pairs,
cooperative transmission by relay nodes has the potential
to improve the overall network performance. In \cite{sadek}, the authors proposed two cooperative cognitive protocols for a multiple access system with a single relay. The relay aids the transmitting nodes transmission during their idle time slots. The secondary throughput of the proposed protocol as well as the delay of symmetric nodes were investigated.

Recently, the idea of emerging cooperative communications and secondary utilization of the spectrum has got a wide attention \cite{simeone,close,khattab,erph,krikidis2009protocol,krikidis2010stability,bao2010stable,Sult1212:Cooperative,krikidis2012stability,wcm,myprotocol}.
In \cite{simeone}, the authors consider a cooperative scheme where the secondary transmitter (ST) is used as a relay node for the undelivered packets of the higher priority user. The authors suggested the use of an admitting parameter to control the relaying fraction. In \cite{close}, an extension of the problem with multiple STs acting as relays for the undelivered packets of the primary user was proposed, with and without opportunistic sensing scheme. In addition the authors of \cite{close} considered priority in transmission is given to the relaying queues. In \cite{khattab}, the authors assumed that the cognitive transmitter will be allowed to use the channel if the primary transmitter (PT) is not using the spectrum. A priority of transmission is given to the relaying packets over the secondary packets. It is assumed that the secondary decides to relay a certain fraction of the undelivered packets of the primary user to minimize the secondary queueing delay subject to a power
budget used for relaying the primary packets. In \cite{erph}, the authors characterized fundamental issues in a shared channel where users have different priority levels. In addition, the authors investigated the stable-throughput region for a two user cognitive shared channel where the primary user has unconditional access to the channel while the secondary user transmits its packets with some adjustable access probability. The channel is modeled as a multipacket reception (MPR) channel. In \cite{krikidis2009protocol}, the authors proposed a cluster of secondary users helping the primary user with a single relaying queue accessible by all the secondary users. In \cite{krikidis2010stability}, the authors considered a network with two primary users and one secondary user relays their undelivered packets in the free time slots. In \cite{bao2010stable}, a multiple primary users and one secondary user capable of relaying is considered. The secondary users are capable of relaying the primary packets. In \cite{Sult1212:Cooperative}, the authors investigated the stability region of a novel multiple channel access protocol for secondary users capable of relaying the undelivered primary packets. Due to queue interaction, the authors provided inner and outer bounds on the stability region.

 In \cite{krikidis2012stability,wcm,myprotocol}, the authors incorporated energy harvesting technologies, where terminals harvest energy from the environment, with cooperative communications. The authors of \cite{krikidis2012stability} investigated the impact of cooperation on the stable throughput of the source in a wireless three-node network topology (source-relay-destination) with energy harvesting
nodes and bursty data traffic and without channel state information (CSI) at the transmitters. In \cite{wcm}, the network model composed of orthogonal channels each owned by a primary user. The secondary user relays the undelivered primary packets and forwards them whenever the primary user is inactive. Inner and outer bounds on the stability region were derived. In \cite{myprotocol}, under the same network model as \cite{wcm} and MPR channel model, the authors investigated the maximum throughput of a new cooperative cognitive protocol for an energy harvesting secondary user cooperating with a primary user.

We consider a slotted time primary and secondary systems. If the primary system
does not have a packet to transmit in a given time slot, then this
time slot is not utilized. These unutilized time slots are wasted
channel resources that can be used by the secondary transmitters and/or receivers to enhance the system performance and spectral efficiency.

 In our work, we propose a cooperative cognitive protocol and characterize its stability region. The cognitive transmitter-receiver pair tries to utilize the periods of silence of the PT in order to increase the reliability of communications against random channel fades for the primary transmissions and to allow the secondary user to utilize the channel effectively. The PT and ST maintain buffers for storing their own data traffic. In addition to its own queue, the ST maintains another queue for storing a fraction of the undelivered primary packets. The secondary receiver (SR) maintains a relaying queue for storing a fraction of the primary undelivered packets. When the secondary system declares empty time slots, the slot is then used to either help the primary system or to allow the secondary packets to be served. For transmission of packets during silence sessions of the PT, we consider two transmission policies that manage the medium access of the ST and the SR. We also investigate two special cases of the proposed system and investigate their stability regions.

We make the following contributions in this paper.
\begin{itemize}
  \item We propose a new cooperative system, which to the
best of our knowledge, has not been proposed before in a
cognitive networks with buffered terminals where the receiver of the cognitive user maintains a data buffer to help the ST to utilize the spectrum via relaying a fraction of the primary undelivered packets.

\item We propose an access probability assigned to each queue of the secondary transmitter-receiver pair and a controllable factor added to each relaying queue. The relaying queues' admitting factors control the arrival processes of the relaying queues and the service process of the primary queue, whereas the access probabilities controls the service processes of the queues.
    \item To manage the access of the ST and the SR, we consider two multiple access policies. Specifically, we investigate the case of random access scheme and time-division multiple-access scheme.
     \item We consider MPR capability added to the primary receiver (PR). Thus, in case of random access scheme adopted by the ST and the SR, the nodes can exploit the MPR capability of the PR when two nodes access at the same time.

    \item     For random access scheme, we provide an inner bound on the stability of the primary-secondary network that is based on the union of two dominant systems. Furthermore, we provide an outer bound on the stability of the primary-secondary network that is based on the intersection of two outer bounds.
        \item We provide two simple systems of the proposed system. Under these systems, the ST is the only node that may cooperate with the PT.
\item We investigate the stability regions of two special cases of the proposed system and derive their exact stability regions.
\item We prove the convexity of the stability regions of the special case systems.

\end{itemize}

The rest of the paper is organized as follows. In the next section,
we describe the system model adopted in this paper. The stable-throughput regions of the proposed systems are considered in Sections \ref{sec1}, \ref{sec2} and \ref{sec3}. In Section \ref{numerical} we provide some numerical results, and finally, conclusions are drawn in Section \ref{conc}.
\section{SYSTEM MODEL} \label{w100}
We consider the cognitive relaying system depicted in
Fig. \ref{term}. We assume that the ST and the SR sense the channel every time slot for $\tau$ seconds to check whether the primary user is idle or not. The sensing process is assumed to be perfect.\footnote{The sensing duration, $\tau$, is assumed to be long enough to make the assumption of perfect sensing valid (see \cite{khattab,krikidis2009protocol} for a similar assumption).} The cognitive system will be able to send a packet each time slot during the idle sessions of the primary user. The main assumptions of the system model at both the MAC and PHY layers are given in this section.
\subsection{PHY Layer Assumptions}
For convenience, we denote the primary transmitter as `${\rm p}$', the primary destination as `${\rm pd}$', the secondary transmitter as `${\rm s}$', and the secondary destination as `${\rm sd}$'. Let $h^t_{\rm j,k}$ denote the channel gain between node ${\rm j}$ and node ${\rm k}$ (${\rm j\rightarrow k}$ link) at instant $t$, where ${\rm j,k} \in \{\rm s,sd,p,pd\}$ and ${\rm j\ne k}$, and it is distributed according to a zero mean circularly symmetric
complex Gaussian random variable with variance $\sigma_{\rm j,k}^2$, i.e., $\mathcal{CN}(0,\sigma_{\rm j,k}^2)$. Channel gains are independent from link to link. Each link is perturbed by complex additive white Gaussian noise (AWGN). The AWGN at receiving node ${\rm k}$ is assumed to be with zero mean and variance $\mathcal{N}_{\rm k}$ Watts. We consider MPR channel model which can
capture the effect of interference and fading at the PHY layer better than the collision channel model \cite{erph}. Packets could survive the interference caused by concurrent
transmissions if the received signal-to-interference-and-noise ratio (SINR) exceeds the threshold required for successful decoding at the receiver. For link ${\rm j}\rightarrow {\rm k}$, the probability of successful reception of the packet sent by node ${\rm j}$ to its receiving node ${\rm k}$ when there is a concurrent transmission from node $\ell$ is given by $\overline{P^\ell_{\rm j,k}}={\rm Pr}\{{\rm SINR}> \gamma^{\left(\rm th\right)}_{\rm j}\}$, where the superscript `$\ell$' denotes the node which causes the interference and $\gamma^{\left(\rm th\right)}_{\rm j}$ denotes the SINR decoding threshold (for details, see Appendix A). The decoding threshold $\gamma^{\left(\rm th\right)}_{\rm j}$ is a function of different factors
in the communication system; it is a function of the application,
the modulation, the signal processing applied at encoder/decoder
sides, error-correction codes, and many other parameters \cite{sadek}. Given the channel model described above, if there is no concurrent transmission, the outage probability between node ${\rm j}$ and node ${\rm k}$ can be calculated as follows:
\begin{equation}\label{145}
\small \begin{split}
{\rm Pr}\{O_{\rm j,k}\}=P_{\rm j,k}&={\rm Pr}\bigg\{|h^t_{\rm j,k}|^2{\mathbb{P}_{\rm j}}<\mathcal{N}_{\rm k}\gamma^{\left(\rm th\right)}_{\rm j}\bigg\}\\&=1-\exp\bigg(-{\frac{\gamma^{\left(\rm th\right)}_{\rm j}\mathcal{N}_{\rm k}}{\sigma^2_{\rm j,k}\mathbb{P}_j}}\bigg)
\normalsize \end{split}
\end{equation}
where $O_{\rm j,k}$ denotes the event that the link ${\rm j}\rightarrow {\rm k}$ is in outage, $\mathbb{P}_{\rm j}$ denotes the transmission power of node ${\rm j}$ in Watts, $\gamma^{\left(\rm th\right)}_{\rm j}=2^{\mathcal{R}_{\rm j}}-1$, $\mathcal{R}_{\rm j}=b/T_{\rm j}/W$, $b$ packets size in bits, $T_{\rm j}$ is the transmission time of node $j$, and $W$ is the channel bandwidth in Hz. Note that the primary user transmits over the whole time slot whenever its queue is nonempty; hence, $T_{\rm p}=T$. On the other hand, both the ST and the SR transmit after sensing the channel for $\tau$ seconds; hence, $T_{\rm s}=T_{\rm sd}=T-\tau$.

From the results in Appendix A, the probability of correct reception of a transmitted packet from node ${\rm j}$ to node ${\rm k}$ when there is a concurrent transmission from node $\ell$ is given by
 \begin{eqnarray} \small\label{193}
 \overline{P^\ell_{\rm j,k}}=\!\frac{\overline{P}_{\rm j,k}}{1+\frac{\mathbb{P}_\ell\gamma^{\left(\rm th\right)}_{\rm j} }{\mathbb{P}_j }\frac{\sigma^2_{\rm \ell,k}}{\sigma_{\rm j,k}^2}}
 \normalfont \end{eqnarray}
where $\overline{\mathcal{X}}=1-\mathcal{X}$.

  For more details regarding the MPR channel model, the reader is referred to \cite{ghez1988stability,ghez1989optimal,erph,naware2005stability,shrader2007random} and the references therein.

\subsection{MAC Layer Assumptions}\label{w111}
We assume that the PT maintains a buffer $Q_{\rm p}$
to store the incoming traffic packets, whereas the ST maintains two buffers: $Q_{\rm s}$ to store its
own arrived traffic packets and $Q_{\rm ps}$ to store a fraction of the undelivered packets of the PT. The SR maintains a relaying queue, denoted by $Q_{\rm sd}$, to store a fraction of the primary undelivered packets. All buffers are assumed to be of
infinite capacity. We consider time-slotted transmissions where all packets have the same size and one time slot is sufficient for the transmission
of a single data packet. The arrival processes of the primary and secondary transmitters are assumed to be independent Bernoulli processes with mean arrival rates $\lambda_{\rm p}$ and $\lambda_{\rm s}$ packet per time slot, respectively. Arrivals to a certain queue are identically distributed from slot to slot. Moreover, arrivals are independent from slot to slot, queue to queue and terminal to terminal.

   \begin{figure}[t]
\normalcolor \centering
  \includegraphics[width=1\columnwidth]{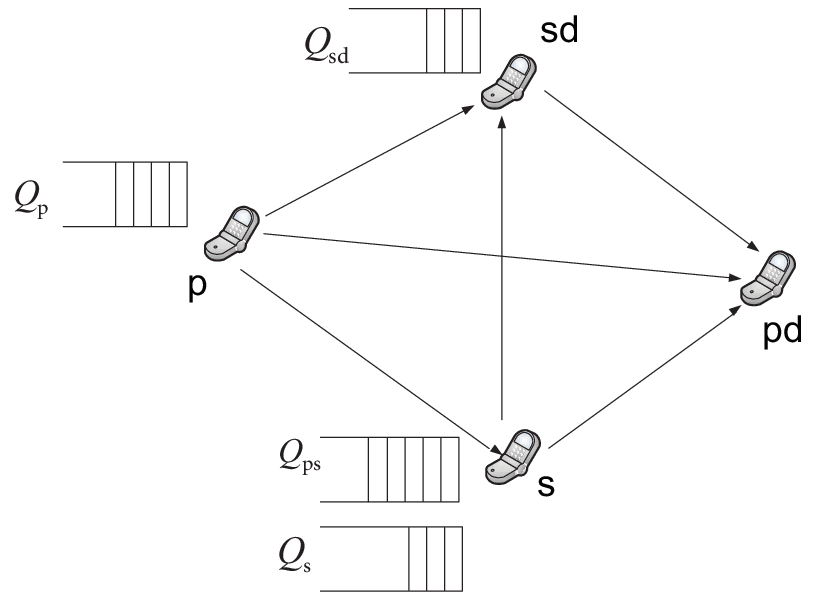}
  \caption{Primary and secondary links of system $\mathcal{S}$. }\label{term}
\end{figure}

The queue size $Q_i^t$, $i\in\{\rm p,s,ps,sd\}$, evolves as follows
\begin{equation}\label{queue}
    Q_i^{t+1}=\bigr(Q_i^t-\mathcal{U}_i^t\bigr)^{+}+\mathcal{A}^t_i
\end{equation}
where $\mathcal{U}_i^t$ is the number of departures in time slot $t$ and $\mathcal{A}^t_i$ denotes the number of arrivals in time slot $t$.
 The function $(.)^{+}$ is defined as $(x)^{+}=\max(x,0)$. We assume that
departures occur before arrivals, and the queue size is measured at the beginning of the time slot \cite{sadek}.

A fundamental performance measure of a communication network is the stability of its queues. We are interested in the queues size. More rigourously, stability can be defined as follows \cite{szpankowski1994stability,sadek}.

\emph{Definition:} Queue $Q_i$, $i\in\{\rm p,s,ps,sd\}$, is stable, if
\begin{equation}\label{stabilityeqn}
   \lim_{\rm t \rightarrow \infty  }{\rm Pr}\{Q_i^t<y\}=F(y) \hbox{ and}  \lim_{\rm y \rightarrow \infty} F(y)=1.
\end{equation}
If the arrival and service processes are strictly stationary, then we can apply Loynes' theorem to check for stability conditions \cite{loynes1962stability,sadek}. This theorem states that if the arrival process and the service process of a queue are strictly stationary processes, and the average service rate is greater than the average arrival rate of the queue, then the queue is stable. If the average service rate is lower than the average arrival rate, then the queue is unstable. Note that this theorem is valid only when queues are decoupled from each other.

 In the proposed system, shown in Fig. \ref{term}, the ST accepts a fraction $f_{\rm s}$ of the undelivered primary packets to be admitted to its relaying queue, whereas the SR accepts a fraction $f_{\rm sd}$ of the undelivered primary packets. We assume that for successfully decoded packets by both the ST and the SR a priority of keeping that packet is one of the optimization parameters of the system, which is denoted by a binary value $\mathcal{P}\in\{0,1\}$. If the priority is assigned to the SR, then $\mathcal{P}=1$; otherwise $\mathcal{P}=0$. To implement this feedback mechanism among different nodes, that possibly receive the same primary packet, we assume that the acknowledgements (ACKs) and negative-acknowledgements (NACKs) messages sent by the node with higher priority of keeping are sent earlier than the messages that are sent by the lower priority node. That is, the node with priority of keeping transmits ACKs and NACKs from $\tau_1<\tau_2$ to $\tau_2$ within the time slot, whereas the other node transmits from $\tau_2<\tau_3$ to $\tau_3$. Note that the primary destination has the highest priority for feedback transmission over both the ST and SR, and it sends the feedback signals over the period $\tau_\circ<\tau_1$ to $\tau_1$. The MAC layer is assumed to obey the following rules.
\begin{itemize}
\item Assign the priority of keeping the undelivered primary packet to the ST or the SR at the beginning of the transmissions.
                                                                 \item The PT transmits the packet at the head of its queue if its queue is nonempty. If the primary queue is empty, the time slot is free.
                                                                 \item If a packet is received successfully by either the PR, the ST, or the SR, the packet is then removed from the PT's queue (the ST or the SR needs to send an ACK if a packet is not decoded correctly by the PR in this case).
                                                                     \item If both the ST and SR decode a packet correctly and the PR cannot decode it, the terminal which has the priority of keeping stores the packet, while the terminal with the lower priority of keeping drops that packet.
                                                                 \item If a packet is not received successfully by the PR, the ST, and the SR, the PT retransmits this packet in the next time slot.
                                                                     \item At each sensed free time slot, the ST and SR may adopt either a random access (RA) scheme or a time-division multiple access (TDMA) scheme. In case of RA scheme, the ST and the SR randomly access the channel (ALOHA random access). The ST transmits a packet from its own queue with probability $\alpha_{\rm s}$, retransmits a packet from the relaying queue with some probability $\alpha_{\rm sp}$, or remains idle with probability $\alpha_i\!=\!1\!-\!\alpha_{\rm s}\!-\!\alpha_{\rm sp}$. The SR retransmits the undelivered packets of the PT with probability $\alpha_{\rm sd}$ or remains idle with probability $1-\alpha_{\rm sd}$. In case of TDMA, the time slots are assigned probabilistically to the ST or the SR. The probability of assigning a time slot to the ST is $\omega$, whereas the probability of assigning a time slot to the SR is $1\!-\!\omega$. Moreover, the ST selects one of its queues for transmission with certain probability. Specifically, the ST selects a packet from its own traffic with probability $\alpha$ or selects a packet from the relaying traffic with probability $1-\alpha$.
                                                                     \item In case of RA scheme, there is a possibility of concurrent transmissions. Packets could survive the interference caused by concurrent transmissions between the ST and the SR, if the received {SINR} exceeds the threshold required for successful decoding at the PR.
                                                               \end{itemize}

It should be pointed out here that the RA-based system can exploit the MPR capability of the PR due to the possibility of concurrent transmissions. This can provide an advantage for the RA-based system over the TDMA-based system at strong MPR capability of the PR. On the other hand, the TDMA-based system can outperform the RA-based system at weak MPR capability because of its collision-free property, which guarantees higher successful transmission probabilities for packets.\footnote{The MPR capability is said to be strong if the receiver is able to decode all concurrent probabilities with successful decoding probability almost equal to the decoding probability when each transmitter communicates with that receiver alone. On the other hand, the MPR capability is said to be weak if the receiver cannot decode any of the transmitters' packets during concurrent transmissions or when the receiver can decode the packets with a very low probability \cite{naware2005stability}.}

We assume that the overhead for transmitting the ACK and NACK messages is very small
compared to packet sizes. The second assumption we make is that the errors and
delay in packet acknowledgement feedback is negligible, which
is reasonable for short length ACK/NACK packets as low rate
codes can be employed in the feedback channel \cite{sadek}. In addition, nodes cannot transmit and receive at the same time. These transmission constraints are common in network systems where terminals are equipped with single transceivers \cite{erph}.

Next, we investigate the stability region of system $\mathcal{S}$ under RA transmission policy. This system is denoted by $\mathcal{S}^{\left(\rm RA\right)}$.
\section{Stability analysis of $\mathcal{S}^{\left(\rm RA\right)}$}\label{sec1}

 The service and arrival processes of the queues are explained as follows. For the primary queue, given that the priority factor $\mathcal{P}=1$, i.e., the priority of keeping the packet is assigned to the SR, a packet can be served if either one of the following events is true: 1) The primary channel is in outage, the SR decides to accept the packet (which occurs with probability $f_{\rm sd}$), and the channel $h^t_{\rm p,sd}$ is not in outage; 2) the primary channel is in outage, the ST decides to accept the packet (which occurs with probability $f_{\rm s}$) and the SR decides not to accept the packet (which occurs with probability $1-f_{\rm sd}$), and the associated link $h^t_{\rm p,s}$ is not in outage; 3) the primary channel is in outage, the ST and the SR both of them decide to accept the primary packet and both of them decode it correctly\footnote{The primary packet will be buffered to the SR queue and dropped from the ST queue due to the priority of keeping assigned to the SR.}; or 4) if the channel between the PT and PR is not in outage, i.e., $\overline {O}^t_{\rm p,pd}$ is true\footnote{$\overline{(.)}$ denotes the complement of the event.}. The service
process can be modeled as
\begin{equation}\label{mu_pp}
    \mathcal{U}_{\rm p}^t=\sum_{\rm m=1}^{4}1\big[A_m^t\big]
\end{equation}
where $1[.]$ denotes the indicator function, and $A_m$, for $m=\{1,2,3,4\}$, are the events described above. From the above argument, it is clear that $ \mathcal{U}_{\rm p}^t$ is stationary process and has a finite mean:
\begin{equation}\label{mu_p}
\small \begin{split}
    \mathcal{E}\{\mathcal{U}_{\rm p}^t\}\!=\!\mu_{\rm p}&\!=\!\overline{P}_{\rm p,pd}\!+\!P_{\rm p,pd}\bigg[f_{\rm sd} \overline{P}_{\rm p,sd}\!+\!(1\!-\! f_{\rm sd}\overline{P}_{\rm p,sd})f_{\rm s} \overline{P}_{\rm p,s} \bigg]
    \normalsize \end{split}
\end{equation}
where $ \mathcal{E}\{.\}$ is the expected value.

If we take the priority of keeping factor into account, the general formula of the average service rate of the PT is given by
\begin{equation}
\small \begin{split}
  \mu_{\rm p}\!=\!\overline{P}_{\rm p,pd}\!& +\!P_{\rm p,pd}\biggr[\mathcal{P}\biggr(f_{\rm sd} \overline{P}_{\rm p,sd} +(1- f_{\rm sd}\overline{P}_{\rm p,sd})f_{\rm s} \overline{P}_{\rm p,s} \biggr)\\& \,\,\,\,\,\,\,\,\,\,\,\,\,\,\,\,\,\,\,\ +\overline{\mathcal{P}}\biggr(f_{\rm s} \overline{P}_{\rm p,s}\!+\!(1\!-\! f_{\rm s}\overline{P}_{\rm p,s})f_{\rm sd} \overline{P}_{\rm p,sd} \!\biggr)\!\biggr]
    \normalsize \end{split}
\end{equation}
  \begin{equation}
  \label{1988}
\small \begin{split}
  \mu_{\rm p}&=\overline{P}_{\rm p,pd}+\underbrace{P_{\rm p,pd}\bigg[f_{\rm sd} \overline{P}_{\rm p,sd}  +f_{\rm s} \overline{P}_{\rm p,s}-f_{\rm s} \overline{P}_{\rm p,s} f_{\rm sd} \overline{P}_{\rm p,sd}\bigg]}_{\mathcal{I}\ge0}.
  \normalsize \end{split}
\end{equation}
It should be mentioned that $\mu_{\rm p}$ is independent of $\mathcal{P}$. We note that without cooperation, the primary mean service rate is $\overline{P}_{\rm p,pd}$. Thus, cooperation increases the primary mean service rate by ${\mathcal{I}}$.

For queue $Q_{\rm s}$, the service process can be modeled as
\begin{equation}\label{mu_s}
\small \begin{split}
    \mathcal{U}_{\rm s}^t&=1\bigg[\{Q^t_{\rm p}=0\} \bigcap A_{\rm s}^t \bigcap \overline {O}^t_{\rm s,sd}\bigcap \{Q^t_{\rm sd}=0\} \bigg]\\& \,\,\,\,\,\,\,\,\,\,\,\,\,\,\,\ +1\bigg[\{Q^t_{\rm p}=0\} \bigcap A_{\rm s}^t \bigcap \overline {O}^t_{\rm s,sd} \bigcap \{Q^t_{\rm sd}\ne0 \}  \bigcap \overline{A}_{\rm sd}\bigg]
    \normalsize \end{split}
\end{equation}
where $\{Q^t_{\rm p}=0\}$ is the event that the primary queue is empty in time slot $t$; $A_{\rm s}^t$ denotes the event that in time slot $t$, the ST assigned the channel to the relaying queue, which occurs with probability
$\alpha_{\rm s}$; $\overline {O}^t_{\rm s,sd}$ denotes the complement of the outage event of the link ${\rm s\rightarrow sd}$; $\{Q^t_{\rm sd}=0\}$ is the event that the SR queue is empty; and $\overline{A}_{\rm sd}$ is the event that the SR is idle. The probability that the primary queue is empty is given by
\begin{equation}\label{empty}
{\rm Pr}\{Q^t_{\rm p}=0\}=1-\frac{\lambda_{\rm p}}{\mu_{\rm p}}.
\end{equation}
From the above argument, and the expression given in $(\ref{empty})$, it is clear that $ \mathcal{U}_{\rm s}^t$ is a stationary process and has a finite mean:
\begin{equation}
\small \begin{split}
    \mathcal{E}\{\mathcal{U}_{\rm s}^t\}&=\mu_{\rm s}\\&=(1-\frac{\lambda_{\rm p}}{\mu_{\rm p}})\overline{P}_{\rm s,sd}\alpha_{\rm s} \bigg[{\rm Pr}\{Q^t_{\rm sd}=0\}+ \overline{\alpha}_{\rm sd} {\rm Pr}\{Q^t_{\rm sd}\ne0\}\bigg].
    \label{mu_ss}
    \normalsize \end{split}
    \end{equation}

Consider now the relaying queue of the ST, $Q_{\rm ps}$. Given that the primary queue is empty in a time slot $t$ and the ST chooses to access the channel using the relaying queue (which occurs with probability $\alpha_{\rm sp}$), a packet from queue $Q_{\rm ps}$ can be served in either one of the following events: 1) If the SR is idle, and the channel between the ST and the PR is not in outage; 2) if the SR does not access the channel (which occurs with probability
$\overline{\alpha}_{\rm sd}$), and its queue is not empty, i.e., $Q^t_{\rm sd} \ne0$, and the channel between the ST and the PR is not in outage; or 3) if the queue $Q_{\rm sd}$ in time slot $t$ is not empty, the SR accesses the channel (which occurs with probability $\alpha_{\rm sd}$), and the complement of the event outage of the link between the ST and the PR, i.e., $1[\overline{O}^t_{\rm s,pd}|\mathcal{T}_{\rm sd}]=1$, where $\overline{O}^t_{j,k}|\mathcal{T}_{\ell}$ denotes the complement of the outage event of the channel between node ${\rm j}$ and node ${\rm k}$ when there is a concurrent transmission by node $\ell$. Mathematically, this can be modeled as follows:
\begin{equation}\label{mu_s}
    \mathcal{\mathcal{U}}_{\rm ps}^t=\sum_{ m=1}^{3}1\big[E_{m}^t\big]
\end{equation}
where $E_m^t$, for $m=\{1,2,3\}$, are the events described above. The expected value of the service process of the queue $Q_{\rm ps}$ is given by
\begin{equation}\label{mu_sssss}
\small \begin{split}
    \mathcal{E}\{\mathcal{U}_{\rm ps}^t\} &=\mu_{\rm ps}\\&=
    \biggr[\bigg({\rm Pr}\{Q^t_{\rm sd}=0\} + \overline{\alpha}_{\rm sd} {\rm Pr}\{Q^t_{\rm sd}\ne0\}\bigg)\overline{P}_{\rm s,pd} \\& \,\,\,\,\,\,\,\,\,\,\,\,\,\,\,\,\,\,\,\,\,\,\,\,\,\,\,\,\,\,\,\ + \alpha_{\rm sd} {\rm Pr}\{Q^t_{\rm sd}\ne0\}\overline{P^{\rm sd}_{\rm s,pd}}\biggr](1-\frac{\lambda_{\rm p}}{\mu_{\rm p}})\alpha_{\rm sp}.
    \normalsize \end{split}
\end{equation}

Consider now the SR's relaying queue $Q_{\rm sd}$. Given that the primary queue is empty in time slot $t$, a packet from queue $Q_{\rm sd}$ can be served if in a time slot $t$ if either one of the following events takes place: 1) If the SR decides to access the channel (which occur with probability $\alpha_{\rm sd})$, the ST has no packets in any of its queues, i.e., $Q_{\rm s}^t=0$ and $Q^t_{\rm ps}=0$, and the complement of the event outage of the link between the SR and the PR; 2) if the SR decides to access the channel, $(Q_{\rm ps}^t\ne0,Q_{\rm s}^t=0)$, the ST does not access the channel (which occurs with probability $\overline{\alpha}_{\rm sp}$), and the link between the SR and the PR is not in outage; 3) if the SR decides to access the channel, the event that the ST's queues are $Q_{\rm ps}^t=0$ and $Q_{\rm s}^t\ne0$, the ST does not access the channel (which occurs with probability $\overline{\alpha}_{\rm s}$), and the link between the SR and the PR is not in outage; 4) if the SR decides to access the channel, the event that the ST's queues are $Q_{\rm ps}^t\ne0$ and $Q_{\rm s}^t\ne0$, the ST does not access the channel (which occurs with probability $\alpha_{\rm i}=1-\alpha_{\rm sp}-\alpha_{\rm s}$), and the link between the SR and the PR is not in outage; 5) if the SR decides to access the channel, $Q_{\rm ps}^t$ is nonempty, the ST accesses the channel (with probability $\alpha_{\rm sp}$), and the complement of the outage event of the link between the SR and the PR given a transmission between the ST and the PR, i.e., $1[\overline{O}^t_{\rm sd,pd}|\mathcal{T}_{\rm ss}]=1$; or 6) if the SR decides to access the channel, $Q_{\rm s}^t$ is nonempty, the ST accesses the channel (which occurs with probability $\alpha_{\rm s}$), and the link between the SR and the PR is not in outage. This can be modeled as:
\begin{equation}\label{mu_s}
    \mathcal{U}_{\rm sd}^t=\sum_{\rm m=1}^{6}1\big[\mathcal{F}^t_m\big]
\end{equation}
where $\mathcal{F}_{m}^t$, for $m=\{1,2,\dots,6\}$, are the events described above. The expected value of the service process of the queue SR is given by

\begin{equation}\label{mu_sd1}
\small
\begin{split} 
    \mathcal{E}\{\mathcal{U}_{\rm sd}^t\}&\!=\! \mu_{\rm sd} \\  &\!=\!(1\!-\!\frac{\lambda_{\rm p}}{\mu_{\rm p}}) \alpha_{\rm sd}\\&  \,\,\,\ \!\times\! \Biggr[\!\overline{P}_{\rm sd,pd}\!\bigg(\!{\rm Pr}\{Q^t_{\rm ps}\!=\!0,Q^t_{\rm s}\!=\!0\}\!+\!\overline{\alpha}_{\rm sp}{\rm Pr}\{Q^t_{\rm ps}\!\ne\!0,Q^t_{\rm s}\!=\!0\} \!\\&  \,\,\,\,\,\,\,\ \!+\!\overline{\alpha}_{\rm s} {\rm Pr}\{Q^t_{\rm ps}\!=\!0,Q^t_{\rm s}\!\ne\!0\} \!+\!\alpha_i {\rm Pr}\{Q^t_{\rm ps}\!\ne\!0,Q^t_{\rm s}\!\ne\!0\}\!\bigg) \!\\& \,\,\,\,\,\,\,\ +\!(\alpha_{\rm sp}{\rm Pr}\{Q^t_{\rm ps}\!\ne\!0\}\! +\!\alpha_{\rm s}{\rm Pr}\{Q^t_{\rm s}\!\ne\!0\}) \overline{P^{\rm s}_{\rm sd,pd}}\!\Biggr].
    \normalsize \end{split}
\end{equation} 

The arrival process to the relaying queue $Q_{\rm ps}$ can be described as follows. Given that $\mathcal{P}=1$, the PT's queue is not empty, i.e., $\{Q_{\rm p}^t>0\}$, the associated channel between PT and PR is in outage, the ST decides to accept the packet, and the channel between the PT and the ST is not in outage, the arrival to $Q_{\rm ps}$ is either one of the following events: 1) The event that the SR decides to accept the packet from the PT, and the associated channel between the PT and the SR is in outage; or 2) if the SR decides not to accept the packet. The process is modeled as
\begin{equation}\label{arrival_ps}
    \mathcal{A}^t_{\rm ps}=\sum_{m=1}^{2}1[\mathcal{W}_m]
\end{equation}
where $\mathcal{W}_m$, $m\in\{1,2\}$, are the events described above, and ${\rm Pr}\{Q_{\rm p}^t>0\}=\frac{\lambda_{\rm p}}{\mu_{\rm p}}$.
The process is stationary and the expected value of the arrival process is expressed as
  \begin{eqnarray} \small\label{mean_arrival_ps}
\lambda_{\rm ps}=\frac{\lambda_{\rm p}}{\mu_{\rm p}}P_{\rm p,pd} \bigr(1- f_{\rm sd}\overline{P}_{\rm p,sd}\bigr) f_{\rm s} \overline{P}_{\rm p,s}.
 \normalfont \end{eqnarray}
Adding the priority factor, the mean arrival rate of the queue $Q_{\rm ps}$ is given by
\begin{eqnarray} \small
  \lambda_{\rm ps}=\frac{\lambda_{\rm p}}{\mu_{\rm p}} P_{\rm p,pd} \biggr(1- \mathcal{P} f_{\rm sd}\overline{P}_{\rm p,sd} \biggr)f_{\rm s}\overline{P}_{\rm p,s}.
 \normalfont \end{eqnarray}

The arrival process to $Q_{\rm sd}$ can be described as follows. The event that the primary has packets, i.e., $\{Q_{\rm p}^t>0\}$, the SR decides to accept a packet from the PT, i.e., $1[W^t_{\rm sd}]=1$, the link ${\rm p\rightarrow sd}$ is not in outage, and the link ${\rm p\rightarrow pd}$ is in outage. The process can be modeled as
\begin{equation}\label{arrival_ps}
    \mathcal{A}^t_{\rm sd}=1\bigg[W^t_{\rm sd}\bigcap \{Q_{\rm p}^t>0\}\bigcap O^t_{\rm p,pd} \bigcap \overline{O}^t_{\rm p,sd}\bigg].
\end{equation}
The process is stationary and the expected value of the arrival process to the queue $Q_{\rm sd}$ is expressed as
  \begin{equation}\label{mean_arrival_ps}
  \lambda_{\rm sd}=f_{\rm sd} P_{\rm p,pd}\overline{P}_{\rm p,sd} \frac{\lambda_{\rm p}}{\mu_{\rm p}}.
\end{equation}

If we involve $\mathcal{P}$, the mean arrival rate of the SR queue is given by
\begin{eqnarray} \small
    \lambda_{\rm sd}=\frac{\lambda_{\rm p}}{\mu_{\rm p}} P_{\rm p,pd} \biggr(1- \overline{\mathcal{P}} f_{\rm s}\overline{P}_{\rm p,s} \biggr)f_{\rm sd}\overline{P}_{\rm p,sd}.
 \normalfont \end{eqnarray}

Since the mean service rates at nodes ${\rm s}$, ${\rm ps}$ and SR depend on each other's queue size, these queues are
called interacting queues, and consequently the rates of the individual
departure processes cannot be computed directly. In order
to overcome this problem, we utilize the idea of stochastic
dominance, which has been applied before to analyze
interacting queues in ALOHA systems \cite{luo1999stability,rao1988stability,erph,sadek}, to obtain inner bounds on the stability region. For the outer bounds, we upper bound the queues service rates such that the service rates of the queues become decoupled.
\subsection{$\mathcal{S}^{\left(\rm RA\right)}$: Inner Bound}
The inner bound is the union over two inner bounds based on two dominant systems.
\subsubsection{First Dominant System }
In this system designated as $\mathcal{S}_1 $, $Q_{\rm sd}$ and $Q_{\rm s}$ send dummy packets when their queues are
empty, and the $Q_{\rm ps}$ behaves exactly as it would in the original system $\mathcal{S}^{\left(\rm RA\right)}$. Now, we can write down the service and arrival rates of the interacting queues, i.e., $Q_{\rm s}$, $Q_{\rm ps}$ and $Q_{\rm sd}$ as follows. The mean service rates of $Q_{\rm s}$ and $Q_{\rm ps}$ are given by
\begin{equation}\label{mu_ss}
    \mu_{\rm s}=(1-\frac{\lambda_{\rm p}}{\mu_{\rm p}})(1-P_{\rm s,sd})\alpha_{\rm s}  \overline{\alpha}_{\rm sd},
    \end{equation}
\begin{eqnarray} \small\label{mu_ssr}
\mu_{\rm ps}&=&(1-\frac{\lambda_{\rm p}}{\mu_{\rm p}})\alpha_{\rm sp}\bigg[ \overline{\alpha}_{\rm sd}\overline{P}_{\rm s,pd} +\alpha_{\rm sd} \overline{P^{\rm sd}_{\rm s,pd}}\bigg].
 \normalfont \end{eqnarray}

The probability ${\rm Pr}\{Q_{\rm ps}=0\}$ is given by
\begin{eqnarray} \small
{\rm Pr}\{Q_{\rm ps}=0\}=1-\frac{\lambda_{\rm ps}}{\mu_{\rm ps}}.
 \normalfont \end{eqnarray}
Therefore,
\begin{equation}
\label{mu_sd}
\small \begin{split}
  \mu_{\rm sd}&\!=\!(1\!-\!\frac{\lambda_{\rm p}}{\mu_{\rm p}}) \alpha_{\rm sd}\biggr[\bigg(\overline{\alpha}_{\rm s}(1\!-\!\frac{\lambda_{\rm ps}}{\mu_{\rm ps}})\!+\!\alpha_i\frac{\lambda_{\rm ps}}{\mu_{\rm ps}}\bigg) \overline{P}_{\rm sd,pd} \!\\&\,\,\,\,\,\,\,\,\,\,\,\,\,\,\,\,\,\,\,\,\,\,\,\,\,\,\,\,\,\,\,\,\,\,\,\,\,\,\,\,\,\,\,\,\,\,\,\,\,\,\,\,\,\,\,\,\,\,\,\,\,\,\,\,\,\,\,\,\,\,\,\,\,\,\,\,\,\ +\!(\alpha_{\rm sp}\frac{\lambda_{\rm ps}}{\mu_{\rm ps}}\!+\!\alpha_{\rm s})\overline{P^{\rm s}_{\rm sd,pd}}\biggr].
  \normalsize \end{split}
\end{equation}

According to the construction of the dominant system $\mathcal{S}_1$, it is easy to see that the queues of the dominant system are
never less than those of the original system, provided they are
both initialized identically (with
the same initial conditions for queue sizes in both the original
and dominant system). This is because, in the dominant
system $\mathcal{S}_1$, the SR transmits dummy packets even if it does not have
any packets in its queue, and therefore interferes with ST in all
cases that it would in the original system. Therefore, if the queues at
all nodes are stable in the dominant system, then the corresponding
queues in the original system must be stable. The first inner bound $R(\mathcal{S}_1)$ which is based on $\mathcal{S}_1$
is given by the closure of the rate pairs $(\lambda_{\rm p},\lambda_{\rm s})$ constrained
by equations shown above as $f_{\rm s}$, $f_{\rm sd}$, $\alpha_{\rm s}$, $\alpha_{\rm sp}$, and $\alpha_{\rm sd}$ vary over $[0,1]$, and $\mathcal{P}$ varies over $\{0,1\}$ \cite{rao1988stability,erph}. For a fixed $\lambda_{\rm p}$, the maximum secondary stable throughput is obtained via solving the following optimization problem (as in \cite{erph,sadek}):
\begin{equation}\label{19099}
\small \begin{split}
   & \underset{\alpha_{\rm s},\alpha_{\rm sp},\alpha_{\rm pd},f_{\rm s},f_{\rm sd},\mathcal{P}}{\max.}  \,\,\,\,\,\,\,\,\,\,\,\,\,\,\,\ \mu_{\rm s} \\ & \,\,\,\,\,\,\,\,\,\,\,\ {\rm s.t.}    \ 0\!\le\! \alpha_{\rm s}, \alpha_{\rm sp} ,\alpha_{\rm sd},f_{\rm s},f_{\rm sd}\!\le\! 1, \ \mathcal{P} \in \{0,1\},\ \alpha_{\rm s}\!+\!\alpha_{\rm sp} \!\le\! 1 \\  & \,\,\,\,\,\,\,\,\,\,\,\,\,\,\,\,\,\,\,\,\,\,\,\,\,\,\,\,\,\,\,\,\,\ \lambda_{\rm p} \!\le\! \mu_{\rm p},\ \lambda_{\rm ps}\!\le\!\mu_{\rm ps}, \ \lambda_{\rm sd}\!\le\!\mu_{\rm sd}.
    \normalsize \end{split}
\end{equation}
\subsubsection{Second Dominant System}
The second dominant system is designated as $\mathcal{S}_2$, where the ST is the one that sends dummy packets from $Q_{\rm s}$ and $Q_{\rm ps}$, i.e., ${\rm Pr}\{Q_{\rm s}\!=\!0\}\!=\!{\rm Pr}\{Q_{\rm ps}\!=\!0\}\!=\!0$, and the SR behaves exactly as it would in the original system $\mathcal{S}^{\left(\rm RA\right)}$. The mean service rate of $Q_{\rm sd}$ is given by
\begin{eqnarray} \small\label{mu_ssss2}
   \mu_{\rm sd}=(1-\frac{\lambda_{\rm p}}{\mu_{\rm p}}) \alpha_{\rm sd}\bigg[\alpha_i \overline{P}_{\rm sd,pd}+(\alpha_{\rm sp}+\alpha_{\rm s})\overline{P^{\rm s}_{\rm sd,pd}}\bigg].
 \normalfont \end{eqnarray}
The probability that $Q_{\rm sd}$ is empty is given by
\begin{eqnarray} \small\label{mu_ss2}
{\rm Pr}\{Q_{\rm sd}=0\}=1-\frac{\lambda_{\rm sd}}{\mu_{\rm sd}}.
     \normalfont \end{eqnarray}
    Thus, the mean service rate of the ST's queues are given by
\begin{eqnarray} \small\label{mu_ss3}
  \mu_{\rm s}&=&(1-\frac{\lambda_{\rm p}}{\mu_{\rm p}})\overline{P}_{\rm s,sd}\ \alpha_{\rm s} \bigg[1- \alpha_{\rm sd}\frac{\lambda_{\rm sd}}{\mu_{\rm sd}}\bigg],
     \normalfont \end{eqnarray}
\begin{equation}\label{mu_sss2}
\small \begin{split}
 \mu_{\rm ps}&=(1-\frac{\lambda_{\rm p}}{\mu_{\rm p}})\alpha_{\rm sp}\bigg[\big(1-\alpha_{\rm sd}\frac{\lambda_{\rm sd}}{\mu_{\rm sd}}\big)\overline{P}_{\rm s,pd}\notag +\alpha_{\rm sd} \frac{\lambda_{\rm sd}}{\mu_{\rm sd}}\overline{P^{\rm sd}_{\rm s,pd}}\bigg].
 \normalsize \end{split}
\end{equation}

The second inner bound for the stable-throughput region of $\mathcal{S}$, $R(\mathcal{S}_2)$, which is based on the dominant system $\mathcal{S}_2$, can be obtained by formulating a constrained optimization problem similar to
that discussed above for the first dominant system, where we fix $\lambda_{\rm p}$ and maximize
$\mu_{\rm s}$ as $f_{\rm s}, f_{\rm ps}, \alpha_{\rm s},\alpha_{\rm sp}$ and $\alpha_{\rm sd}$ vary over $[0,1]$ and $\mathcal{P}$ varies over $\{0,1\}$.

\subsection{$\mathcal{S}^{\left(\rm RA\right)}$: Outer Bound}
Here we provide two outer bounds for $\mathcal{S}^{\left(\rm RA\right)}$.
\subsubsection{First Outer Bound}
The first outer bound for the $\mathcal{S}^{\left(\rm RA\right)}$, denoted by $\mathcal{S}_1^{\left(o\right)}$, can be obtained by upper bounding the joint probability identities and using Bayes' theorem \cite{Sult1212:Cooperative}. More specifically,
\begin{equation}
\small \begin{split}
{\rm Pr}\{\!Q^t_{\rm sd}\!=\!0\}\!+\! \overline{\alpha}_{\rm sd} {\rm Pr}\{\!Q^t_{\rm sd}\!\ne\!0\} &\!\le\! {\rm Pr}\{Q^t_{\rm sd}\!=\!0\}\!+\!{\rm Pr}\{\!Q^t_{\rm sd}\!\ne\!0\}\!=\!1,
 \normalsize \end{split}
\end{equation}
 and
\begin{equation}
\small \begin{split}
 \bigg({\rm Pr}\{Q^t_{\rm sd}=0\} \!+\!& \overline{\alpha}_{\rm sd} {\rm Pr}\{Q^t_{\rm sd}\ne0\}\bigg)\overline{P}_{\rm s,pd}\notag \\& +\alpha_{\rm sd} {\rm Pr}\{Q^t_{\rm sd}\ne0\}\overline{P^{\rm sd}_{\rm s,pd}} \le \overline{P}_{\rm s,pd}.
  \normalsize \end{split}
\end{equation}
Based on Bayes' theorem, we have
\begin{equation}\label{mu_rr}
\small \begin{split}
{\rm Pr}\{a,\mathcal{B}\}&={\rm Pr}\{a|B\} {\rm Pr}\{\mathcal{B}\} \le {\rm Pr}\{\mathcal{B}\} \ \ \ \\& \hbox{or} \\ \ \ \ {\rm Pr}\{a,\mathcal{B}\}&={\rm Pr}\{\mathcal{B}|a\} {\rm Pr}\{a\} \le {\rm Pr}\{a\}
  \normalsize \end{split}
\end{equation}
where $a$ and $\mathcal{B}$ are any two arbitrary events. We can upper bound the following quantities in formula (\ref{mu_sd1}):
\begin{equation}\label{mu_sss}
\small \begin{split}
{\rm Pr}\{Q^t_{\rm ps}=0,Q^t_{\rm s}=0\}&\le {\rm Pr}\{Q^t_{\rm ps}=0\},\\ {\rm Pr}\{Q^t_{\rm ps}\ne0,Q^t_{\rm s}=0\}&\le  {\rm Pr}\{Q^t_{\rm ps}\ne0\},\\ \ {\rm Pr}\{Q^t_{\rm ps}=0,Q^t_{\rm s}\ne0\} &\le {\rm Pr}\{Q^t_{\rm ps}=0\},\\  {\rm Pr}\{Q^t_{\rm ps}\ne0,Q^t_{\rm s}\ne0\}&\le {\rm Pr}\{Q^t_{\rm ps}\ne0\}.
    \normalsize \end{split}
\end{equation}

Based on the above facts, the mean service rates of the ST's queues can be upper bounded as follows:
\begin{equation}
    \mu_{\rm s}\le(1-\frac{\lambda_{\rm p}}{\mu_{\rm p}})\overline{P}_{\rm s,sd}\alpha_{\rm s}, \,\,\ \mu_{\rm ps}\le(1-\frac{\lambda_{\rm p}}{\mu_{\rm p}})\alpha_{\rm sp}\overline{P}_{\rm s,pd}.
    \end{equation}
Therefore, the mean service rate of $Q_{\rm sd}$ is upper bounded as follows:
\begin{equation}\label{mu_sg}
\small \begin{split}
  \mu_{\rm sd}&\le (1\!-\!\frac{\lambda_{\rm p}}{\mu_{\rm p}}) \alpha_{\rm sd}\biggr[\bigg((1\!-\!\frac{\lambda_{\rm ps}}{\mu_{\rm ps}}) +\!\overline{\alpha}_{\rm sp}\frac{\lambda_{\rm ps}}{\mu_{\rm ps}}\! +\!\overline{\alpha}_{\rm s}(1\!-\!\frac{\lambda_{\rm ps}}{\mu_{\rm ps}})\!\\&\,\,\,\,\,\,\,\,\,\,\,\,\,\,\,\,\,\,\,\,\,\,\,\,\,\,\,\,\,\,\ +\!{\alpha}_i\frac{\lambda_{\rm ps}}{\mu_{\rm ps}}\bigg) \overline{P}_{\rm sd,pd} \! +\!\bigg(\alpha_{\rm sp}\frac{\lambda_{\rm ps}}{\mu_{\rm ps}}\!+\!\alpha_{\rm s}\bigg)\overline{P^{\rm s}_{\rm sd,pd}}\!\biggr].
  \normalsize \end{split}
\end{equation}
When the inequalities hold to equalities, the queues are not interacting anymore and therefore we can obtain the outer bound by solving a constrained optimization problem to get the closure $(\lambda_{\rm p},\lambda_{\rm s})$. The optimization problem is similar to (\ref{19099}).

The optimization problems of the first and second dominant systems and the first outer bound are solved numerically using MatLab's fmincon \cite{6636823,Sult1210:Optimal,4472701,6177245,6568963}. Since the problems are nonconvex, the solver produces a locally optimum solution. To
increase the likelihood of obtaining the global optimum, the program is run
many times, say $10000$ times, with different initializations of the optimization variables.
\subsubsection{Second Outer Bound}
Another outer bound which can be stated analytically is obtained as follows. Using (\ref{1988}),
  \begin{equation}
  \label{19882}
\small \begin{split}
  \mu_{\rm p}&\!=\!\overline{P}_{\rm p,pd}\!+\!P_{\rm p,pd}\bigg[f_{\rm sd} \overline{P}_{\rm p,sd}  \!+\!f_{\rm s} \overline{P}_{\rm p,s}\!-\!f_{\rm s} \overline{P}_{\rm p,s} f_{\rm sd} \overline{P}_{\rm p,sd}\bigg] \\& \!\le\! \overline{P}_{\rm p,pd}\!+\!P_{\rm p,pd} \bigg(\!1\!-\! P_{\rm p,sd} P_{\rm p,s}\!\bigg)\!=\!1\!-\!P_{\rm p,pd} P_{\rm p,sd} P_{\rm p,s},
  \normalsize \end{split}
\end{equation}

\begin{equation}
    \mu_{\rm s}\le(1-\frac{\lambda_{\rm p}}{\mu_{\rm p}})\overline{P}_{\rm s,sd}\alpha_{\rm s}\le (1-\frac{\lambda_{\rm p}}{\mu_{\rm p}})\overline{P}_{\rm s,sd}.
    \end{equation}
 When the inequality (\ref{19882}) holds to equality and applying Loynes' theorem, we get
    \begin{equation}
    \label{toko2}
    \lambda_{\rm s}< \mu_{\rm s}\le (1-\frac{\lambda_{\rm p}}{ 1\!-\!P_{\rm p,pd} P_{\rm p,sd} P_{\rm p,s} })\overline{P}_{\rm s,sd}.
    \end{equation}

    Denote the second outer bound as $\mathcal{S}_2^{\left(o\right)}$. The outer bound can be characterized by the rate pairs
    \begin{equation}
\small \begin{split}
   R(\mathcal{S}_2^{\left(o\right)})&\!=\!\biggr\{ (\lambda_{\rm p},\lambda_{\rm s})\!:\!  \frac{\lambda_{\rm s}}{\overline{P}_{\rm s,sd}}\!+\!\frac{\lambda_{\rm p}}{ 1\!-\!P_{\rm p,pd} P_{\rm p,sd} P_{\rm p,s}}\!<\! 1\biggr\}.
   \normalsize \end{split}
\end{equation}

The outer bound, $\mathcal{S}^{\left(o\right)}$, of $\mathcal{S}^{\left(\rm RA\right)}$ is the intersection of the two outer bounds, i.e., $R(\mathcal{S}^{\left(o\right)})=R(\mathcal{S}_1^{\left(o\right)})\bigcap R(\mathcal{S}_2^{\left(o\right)})$.
Note that since the service rates of the queues in $\mathcal{S}_1^{\left(o\right)}$ are upper bounded to obtain the mean service rates of the queues in $\mathcal{S}_2^{\left(o\right)}$ (see (\ref{19882}) to (\ref{toko2})), $R(\mathcal{S}_1^{\left(o\right)})$ is contained inside $R(\mathcal{S}_2^{\left(o\right)})$, i.e., $R(\mathcal{S}_1^{\left(o\right)})\bigcap R(\mathcal{S}_2^{\left(o\right)})=R(\mathcal{S}_1^{\left(o\right)})$.

{\it Note that any point (rate pair $(\lambda_{\rm p},\lambda_{\rm s})$) below the outer bound is either stable or unstable, but all points above the outer bound are unstable. On the other hand, all points below the the inner bound are stable, but any point above the inner bound is either stable or unstable.}

\subsection{The case of strong MPR}
In this case, the PR can decode the packet sent by the ST and the SR with the same probability as in case of no interference. That is, $\overline{P^{\rm sd}_{\rm s,pd}}=\overline{P}_{\rm s,pd}$ and $\overline{P^{\rm s}_{\rm sd,pd}}=\overline{P}_{\rm sd,pd}$. In such case, the ST and the SR can transmit at the same time without any losses for their queues mean service rates with respect to the interference-free access. This case can happen if $\gamma^{(\rm th)}_{\rm s}=2^{b/(TW(1-\tau/T))}-1$ is much less than $1$, i.e., when the packet size $b$ is much smaller than the product $TW$. The service and arrival rates of the queues are given by
  \begin{equation}
  \label{1988}
\small \begin{split}
  \mu_{\rm p}&=\overline{P}_{\rm p,pd}+P_{\rm p,pd}\bigg[f_{\rm sd} \overline{P}_{\rm p,sd}  +f_{\rm s} \overline{P}_{\rm p,s}-f_{\rm s} \overline{P}_{\rm p,s} f_{\rm sd} \overline{P}_{\rm p,sd}\bigg],\\
  \mu_{\rm s}&=(1-\frac{\lambda_{\rm p}}{\mu_{\rm p}})\overline{P}_{\rm s,sd},\\
    \mu_{\rm sd} & \!=\!(1\!-\!\frac{\lambda_{\rm p}}{\mu_{\rm p}}) \overline{P}_{\rm sd,pd},\\
  \mu_{\rm ps}&=(1-\frac{\lambda_{\rm p}}{\mu_{\rm p}})  \overline{{P}}_{\rm s,pd},\\
  \lambda_{\rm ps}&=\frac{\lambda_{\rm p}}{\mu_{\rm p}} P_{\rm p,pd} \biggr(1- \mathcal{P} f_{\rm sd}\overline{P}_{\rm p,sd} \biggr)f_{\rm s}\overline{P}_{\rm p,s},\\
    \lambda_{\rm sd}&=\frac{\lambda_{\rm p}}{\mu_{\rm p}} P_{\rm p,pd} \biggr(1- \overline{\mathcal{P}} f_{\rm s}\overline{P}_{\rm p,s} \biggr)f_{\rm sd}\overline{P}_{\rm p,sd}.
    \normalsize \end{split}
\end{equation}

The optimal pair of fractions ($f_{\rm s}, f_{\rm sd}$) and $\mathcal{P}$ can be obtained via finding the set of points that satisfies both the relaying queues stability constraints. That is,

\begin{equation}
\small \begin{split}
\lambda_{\rm ps}\le  \mu_{\rm ps} \Leftrightarrow    \frac{\lambda_{\rm p}}{\mu_{\rm p}} P_{\rm p,pd} \overline{ \mathcal{P} f_{\rm sd}\overline{P}_{\rm p,sd} }f_{\rm s}\overline{P}_{\rm p,s}\le (1-\frac{\lambda_{\rm p}}{\mu_{\rm p}})  \overline{{P}}_{\rm s,pd}
    \normalsize \end{split}
\end{equation}

\begin{eqnarray} \small
   \lambda_{\rm sd}\!\le\!  \mu_{\rm sd}  \Leftrightarrow \frac{\lambda_{\rm p}}{\mu_{\rm p}} P_{\rm p,pd} \overline{ \overline{\mathcal{P}} f_{\rm s}\overline{P}_{\rm p,s}}f_{\rm sd}\overline{P}_{\rm p,sd}\!\le\! (1\!-\!\frac{\lambda_{\rm p}}{\mu_{\rm p}}) \overline{P}_{\rm sd,pd}.
 \normalfont \end{eqnarray}

We emphasize here the following. The RA-based system $\mathcal{S}^{\left(\rm RA\right)}$ requires less cooperation between the ST and the SR for its implementation relative to the TDMA-based system $\mathcal{S}^{\left(\rm TDMA\right)}$.
\section{$\mathcal{S}$ under TDMA: $\mathcal{S}^{\left(\rm TDMA\right)}$} \label{sec2}
Under TDMA transmission policy, the sensed free time slots are shared among the ST and the SR probabilistically. The probability of assigning a time slot to ST is $\omega$, whereas the probability of assigning a time slot to SR is $1-\omega$. Moreover, the ST selects one of its queues for transmission randomly whenever it gets a free time slot. The probability that the ST selects $Q_{\rm s}$ is $\alpha$, whereas the probability that the ST selects $Q_{\rm ps}$ is $1-\alpha$. The mean service rate of the primary queue and the arrival rates to the relaying queues are given by the expressions in the previous Section. The service rate of the relaying queue $Q_{\rm sd}$ under TDMA policy is given by
\begin{equation}\label{mu_sdtdma}
\small \begin{split}
    \mu_{\rm sd} \!=\!(1\!-\!\frac{\lambda_{\rm p}}{\mu_{\rm p}})(1-\omega) \overline{P}_{\rm sd,pd}\!
    \normalsize \end{split}
\end{equation}
where $1-\omega$ is the probability of assigning the current time slot to the SR for transmission. The expression of $\mu_{\rm sd}$, Eqn. (\ref{mu_sdtdma}), is explained as follows. A packet at the head of $Q_{\rm sd}$ is served if the primary queue is empty, which occurs with probability $1-\lambda_{\rm p}/\mu_{\rm p}$; the time slot is assigned to the SR, which occurs with probability $1-\omega$; and the link ${\rm sd\rightarrow pd}$ is not in outage.

In a similar fashion, the mean service rates of the secondary own and relaying queues are given by
\begin{equation}\label{mu_stdma}
\small \begin{split}
    \mu_{\rm s} \!=\!(1\!-\!\frac{\lambda_{\rm p}}{\mu_{\rm p}})\omega \alpha \overline{P}_{\rm s,sd}\!
    \normalsize \end{split}
\end{equation}
\begin{equation}\label{mu_pstdma}
\small \begin{split}
    \mu_{\rm ps} \!=\!(1\!-\!\frac{\lambda_{\rm p}}{\mu_{\rm p}})\omega (1-\alpha) \overline{P}_{\rm s,pd}\!
    \normalsize \end{split}
\end{equation}
where $\omega$ is the probability of assigning the current time slot to ST and $\alpha$ is the probability that the ST selects the secondary queue $Q_{\rm s}$ for transmission.

The optimization problem which obtains the maximum secondary stable throughput is given by
\begin{equation}\label{19099rr}
\small \begin{split}
   & \underset{\alpha,\omega,f_{\rm s},f_{\rm sd},\mathcal{P}}{\max.}  \,\,\,\,\,\,\,\,\,\,\,\,\,\,\,\ \mu_{\rm s}=(1\!-\!\frac{\lambda_{\rm p}}{\mu_{\rm p}})\omega \alpha \overline{P}_{\rm s,sd} \\ & \,\,\,\,\,\,\,\,\,\,\,\ {\rm s.t.}    \ 0\!\le\! \alpha, \omega ,f_{\rm s},f_{\rm sd}\!\le\! 1, \ \mathcal{P} \in \{0,1\},\ \\  & \,\,\,\,\,\,\,\,\,\,\,\,\,\,\,\,\,\,\,\,\,\,\,\,\,\,\,\,\,\,\,\,\,\ \lambda_{\rm p} \!\le\! \mu_{\rm p},\ \lambda_{\rm ps}\!\le\!\mu_{\rm ps}, \ \lambda_{\rm sd}\!\le\!\mu_{\rm sd}.
    \normalsize \end{split}
\end{equation}
We maximize the secondary mean service rate under the stability of all other queues in the system and as the optimization parameters ${\alpha,\omega,f_{\rm s},f_{\rm sd}}$ and $\mathcal{P}$ vary over their domains.

For a given $f_{\rm s}\in[0,1]$, $f_{\rm sd}\in[0,1]$ and $\mathcal{P}\in\{0,1\}$, the optimization problem (\ref{19099rr}) is a linear program. Since $1~-~\omega$, $\alpha \omega$ and $(1-\alpha)\omega$ sum up to $1$, the optimal parameters are
\begin{equation}\label{mu_pstdma}
\small \begin{split}
  \omega (1-\alpha)=\frac{\frac{\lambda_{\rm p}}{\mu_{\rm p}} P_{\rm p,pd} \biggr(1- \mathcal{P} f_{\rm sd}\overline{P}_{\rm p,sd} \biggr)f_{\rm s}\overline{P}_{\rm p,s} }{(1\!-\!\frac{\lambda_{\rm p}}{\mu_{\rm p}}) \overline{P}_{\rm s,pd}}
    \normalsize \end{split}
\end{equation}
\begin{equation}\label{mu_pstdma}
\small \begin{split}
  1-\omega=\frac{\frac{\lambda_{\rm p}}{\mu_{\rm p}} P_{\rm p,pd} \biggr(1- \overline{\mathcal{P}} f_{\rm s}\overline{P}_{\rm p,s} \biggr)f_{\rm sd}\overline{P}_{\rm p,sd}}{(1\!-\!\frac{\lambda_{\rm p}}{\mu_{\rm p}}) \overline{P}_{\rm sd,pd}}
    \normalsize \end{split}
\end{equation}
\begin{equation}\label{mu_pstdma}
\small \begin{split}
 \omega \alpha=\frac{\lambda_{\rm s}}{(1\!-\!\frac{\lambda_{\rm p}}{\mu_{\rm p}}) \overline{P}_{\rm s,sd}}.
    \normalsize \end{split}
\end{equation}
The optimal solution must satisfy the constraint that $\alpha\omega+\omega(1-\alpha)+(1-\omega)=1$. Hence, the stability region for a fixed $f_{\rm s}\in[0,1]$, $f_{\rm sd}\in[0,1]$ and $\mathcal{P}\in\{0,1\}$ is given in Eqn. (\ref{fgc}) at the top of the following page.
\begin{table*}
\begin{equation}
\small \begin{split}
\label{fgc}
   R(\mathcal{S}^{\left(\rm TDMA\right)})&=\bigg\{(\lambda_{\rm p},\lambda_{\rm s}):\frac{\lambda_{\rm s}}{(1\!-\!\frac{\lambda_{\rm p}}{\mu_{\rm p}}) \overline{P}_{\rm s,sd}}+\frac{\frac{\lambda_{\rm p}}{\mu_{\rm p}} P_{\rm p,pd} \biggr(1- \overline{\mathcal{P}} f_{\rm s}\overline{P}_{\rm p,s} \biggr)f_{\rm sd}\overline{P}_{\rm p,sd}}{(1\!-\!\frac{\lambda_{\rm p}}{\mu_{\rm p}}) \overline{P}_{\rm sd,pd}}+\frac{\frac{\lambda_{\rm p}}{\mu_{\rm p}} P_{\rm p,pd} \biggr(1- \mathcal{P} f_{\rm sd}\overline{P}_{\rm p,sd} \biggr)f_{\rm s}\overline{P}_{\rm p,s} }{(1\!-\!\frac{\lambda_{\rm p}}{\mu_{\rm p}}) \overline{P}_{\rm s,pd}}\le 1\bigg\}.
   \\\\
   \hline
   \normalsize \end{split}
\end{equation}
\end{table*}
\section{The case of $f_{\rm sd}=0$ with and without Prioritized relaying} \label{sec3}
\subsection{$f_{\rm sd}=0$ with Prioritized Relaying} \label{lemma}
In this subsection, we investigate the stability region of the proposed system when $f_{\rm sd}=0$ and with a priority of transmission assigned to the relaying packets over the secondary packets. Under this setting, the ST is the only cooperative terminal with the PT in the network. The ST uses the idle time slots of the PT to transmit a packet from the relaying queue with probability $1$, if the relaying queue is nonempty. If both the primary and relaying queues are empty, the ST transmits a packet from its own queue. The service and arrival rates of each queue in the system is given by
\begin{eqnarray} \small
 \mu_{\rm p} &=& \overline{P}_{\rm p,pd}+f_{\rm s} P_{\rm p,pd}\overline{P}_{\rm p,s}\\
   \lambda_{\rm ps}&=&f _{\rm s} P_{\rm p,pd}\overline{P}_{\rm p,s}\frac{\lambda_{\rm p}}{\mu_{\rm p}} \\
  \mu_{\rm ps} &=& (1-\frac{\lambda_{\rm p}}{\mu_{\rm p}})\overline{P}_{\rm s,pd}\\
  \mu_{\rm s} &=& (1-\frac{\lambda_{\rm p}}{\mu_{\rm p}})(1-\frac{\lambda_{\rm ps}}{\mu_{\rm ps}})\overline{P}_{\rm s,sd}.
 \normalfont \end{eqnarray}
Note that $\mu_{\rm s}$ is proportionally increasing with $(1-\frac{\lambda_{\rm ps}}{\mu_{\rm ps}})$. The term $(1-\frac{\lambda_{\rm ps}}{\mu_{\rm ps}})$ indicates the priority of transmission assigned to the relaying packets over the secondary packets.

Stable-throughput region of the system can be obtained by formulating a constrained optimization problem similar to
that discussed above for the $\mathcal{S}^{\left(\rm RA\right)}$ system, where we fix $\lambda_{\rm p}$ and maximize
$\mu_{\rm s}$ as $f_{\rm s}$ varies over $[0,1]$. That is,

\begin{eqnarray} \small\label{1900099}
    \underset{ f_{\rm s}}{\max.} & &(1-\frac{\lambda_{\rm p}}{\mu_{\rm p}})(1-\frac{\lambda_{\rm ps}}{\mu_{\rm ps}})\overline{P}_{\rm s,sd} \notag \\ {\rm s.t.} & &     \lambda_{\rm p} \le \mu_{\rm p} \notag  \\& &0\le f_{\rm s}\le 1 \notag  \\& & \lambda_{\rm ps}\le \mu_{\rm ps}.
 \normalfont \end{eqnarray}
After some mathematical manipulations, the optimization problem reduces to:
\begin{eqnarray} \small\label{1900099}
    \underset{ f_{\rm s}}{\min.} & &\frac{f_{\rm s} }{\overline{P}_{\rm p,pd}-\overline{P}_{\rm s,pd}}
    \notag \\ {\rm s.t.} & &     f_{\rm s} \ge \frac{\lambda_{\rm p}-\overline{P}_{\rm p,pd}}{K} \notag\\& &0\le f_{\rm s}\le 1 \notag \\& &    f_{\rm s} ({\lambda_{\rm p}}- \overline{P}_{\rm s,pd})\le (\overline{P}_{\rm p,pd}-{\lambda_{\rm p}}{})\frac{\overline{P}_{\rm s,pd}}{K}
 \normalfont \end{eqnarray}
where $K={P_{\rm p,pd}\overline{P}_{\rm p,s}}$. The objective and the constraints of the above optimization problem
are linear; hence, the optimization problem is a linear program.
It can be noted that if $\overline{P}_{\rm s,pd} = \overline{P}_{\rm p,pd}$, the above optimization problem will be reduced to a feasibility problem \cite{boyed}. The relaying queue constraint can be rewritten as
\begin{eqnarray} \small
\lambda_{\rm p}\le  \mu_{\rm p} \frac{\overline{P}_{\rm s,pd}}{f _{\rm s} P_{\rm p,pd}\overline{P}_{\rm p,s} +\overline{P}_{\rm s,pd}} \le \mu_{\rm p}.
\label{ghj}
 \normalfont \end{eqnarray}
The term $\frac{\overline{P}_{\rm s,pd}}{f_{\rm s} P_{\rm p,pd}\overline{P}_{\rm p,s} +\overline{P}_{\rm s,pd}}$ is obviously less than the unity. Based on (\ref{ghj}), the constraint on the relaying queue subsumes that of the primary queue.

The optimal value of the acceptance factor $f_{\rm s}$, according to the quality of the links ${\rm p \rightarrow pd}$ and ${\rm s \rightarrow pd}$, is given by
         \begin{itemize}
\item If $\overline{P}_{\rm p,pd}<\overline{P}_{\rm s,pd}$,
   \[ f^*_{\rm s} = \left\{ \begin{array}{ll}
   [0,1]& \mbox{if $\lambda_{\rm p} =0$} ;\\
         1 & \mbox{if $0<\lambda_{\rm p} \le \overline{P}_{\rm s,pd}$}.\end{array} \right. \]

        \item If $\overline{P}_{\rm p,pd}>\overline{P}_{\rm s,pd}$,
   \[ f^*_{\rm s} = \left\{ \begin{array}{ll}
   [0,1]& \mbox{if $\lambda_{\rm p} =0$} ;\\
         0 & \mbox{if $0<\lambda_{\rm p} \le \overline{P}_{\rm p,pd}$}.\end{array} \right. \]

               \item If $\overline{P}_{\rm p,pd}=\overline{P}_{\rm s,pd}$, the problem is reduced to a feasibility problem. The optimal $f_{\rm s}$ is given by

         \begin{equation}
\small \begin{split}
 f_{\rm s}^*=[0,1].
   \normalsize \end{split}
\end{equation}

         \end{itemize}

         Thus, the ST according to the \textbf{channels quality} chooses the optimal value of the admitting factor $f_{\rm s}$. These conditions have the following intuitive explanation, if on the average, the ${\rm s \rightarrow pd}$ channel is worse than the ${\rm p \rightarrow pd}$ channel, then it is better for the PT to transmit its own packets.

         If the links ${\rm p \rightarrow pd}$ and ${\rm s \rightarrow pd}$ have the same quality, the stability region of the network will not depend on $f_{\rm s}$; hence, setting $f_{\rm s}$ to any value will not change the stability region. However, we would emphasize the following, if we design an energy efficient scheme for the primary user, then setting $f_{\rm s}$ to unity, i.e., $f_{\rm s}=1$, would be the optimal solution. This is because the undelivered primary packet will be delivered to the PR without further energy from the PT, if the ST could decode it. On the other hand, if we design an energy efficient scheme for the secondary user, then setting $f_{\rm s}$ to zero would be the optimal solution. This is because the PT will retransmit the undelivered packets without any aid from the ST; hence, the ST will not spend any energy to deliver those packets. Note that in both cases, we get the exact same stability region, however, the value of $f_{\rm s}$ manages the transmit energy that will be used by a terminal, on the average, to achieve certain energy constraints or requirements. However, this is out of scope of this paper.

The stability region of $\hat{\mathcal{S}}^{(\rm P)}$ is given by Eqn. (\ref{10002}) at the top of this page.

\begin{table*}
\begin{equation}
\small \begin{split}
   R(\hat{\mathcal{S}}^{(\rm P)})&=\bigg\{(\lambda_{\rm p},\lambda_{\rm s}):\lambda_{\rm s}<\frac{1}{\overline{P}_{\rm p,pd}+f_{\rm s}^* K} (\overline{P}_{\rm p,pd}+f_{\rm s}^* K-\frac{K}{\overline{P}_{\rm s,pd}}{f^*_{\rm s} {\lambda_{\rm p}}}-\lambda_{\rm p})\overline{P}_{\rm s,sd}\bigg\}.
   \label{10002}
   \normalsize \end{split}
\end{equation}
\end{table*}

\noindent  Using the optimal value of $f_{\rm s}$ and the stability region equation (\ref{10002}), we can show that the stability region of $\hat{\mathcal{S}}^{(\rm P)}$ is {\bf convex}, specifically the stability region is a {\it polyhedron}. That is, if $\overline{P}_{\rm p,pd}<\overline{P}_{\rm s,pd}$, then $f_{\rm s}^*=1$. The stability region, after some simplifications, is then given by Eqn. (\ref{por}) at the top of the following page.

\begin{table*}
\begin{equation}
\small \begin{split}
\label{por}
   R(\hat{\mathcal{S}}^{(\rm P)})&=\bigg\{(\lambda_{\rm p},\lambda_{\rm s}):\lambda_{\rm s}<\frac{1}{\overline{P}_{\rm p,pd}+K} \Big(\overline{P}_{\rm p,pd}+ K-(1+\frac{K}{\overline{P}_{\rm s,pd}}){ {\lambda_{\rm p}}}\Big)\overline{P}_{\rm s,sd}\bigg\}.
     \\ \\
   \hline
   \normalsize \end{split}
\end{equation}
\end{table*}

If $\overline{P}_{\rm p,pd}>\overline{P}_{\rm s,pd}$, then $f_{\rm s}^*=0$. The stability region, after some simplifications, is then given by

\begin{equation}
\small \begin{split}
   R(\hat{\mathcal{S}}^{(\rm P)})&=\bigg\{(\lambda_{\rm p},\lambda_{\rm s}):\lambda_{\rm s}<(1-\frac{\lambda_{\rm p}}{\overline{P}_{\rm p,pd}})\overline{P}_{\rm s,sd}\bigg\}.
   \label{df}
   \normalsize \end{split}
\end{equation}
The convexity of the stability region of $\hat{\mathcal{S}}^{(\rm P)}$ implies that for any given two stable rate pairs $(\lambda_{\rm p},\lambda_{\rm s})$, the line segment connecting them is also in the set and, hence, is composed of stable rate pairs. Note that based on the stability regions, the envelop of the stability region linearly decreases with $\lambda_{\rm p}$. The degradation rate with increasing $\lambda_{\rm p}$ is given by
   \[ \frac{\partial \lambda_{\rm s}}{\partial \lambda_{\rm p}}= \left\{ \begin{array}{ll}
   -\frac{(\overline{P}_{\rm s,pd}+{K})}{\overline{P}_{\rm p,pd}+K} & \mbox{if $\overline{P}_{\rm p,pd}<\overline{P}_{\rm s,pd}$} ;\\
         -\frac{\overline{P}_{\rm s,sd}}{\overline{P}_{\rm p,pd}}  & \mbox{if $\overline{P}_{\rm p,pd}>\overline{P}_{\rm s,pd}$}.
         \end{array} \right. \]

\subsection{$f_{\rm sd}=0$ without Prioritized Relaying}
 In this subsection, we investigate the stability region of a cooperative cognitive transmitter with no priority assigned to the relaying queue and with adaptive acceptance factor of the primary undelivered packets. The ST uses the idle time slots of the primary user to transmit a packet from the relaying traffic with probability $\alpha_{\rm sp}$ or to transmit a packet from its own traffic with probability $\alpha_{\rm s}=1-\alpha_{\rm sp}$. Setting $f_{\rm sd}=0$ in $\mathcal{S}$, the arrival and service rates of each queue are given by
\begin{eqnarray} \small
 \mu_{\rm p} &=& \overline{P}_{\rm p,pd}+f_{\rm s} P_{\rm p,pd}\overline{P}_{\rm p,s},\\
   \lambda_{\rm ps}&=&f _{\rm s} P_{\rm p,pd}\overline{P}_{\rm p,s}\frac{\lambda_{\rm p}}{\mu_{\rm p}}, \\
  \mu_{\rm ps} &=& \alpha_{\rm sp}(1-\frac{\lambda_{\rm p}}{\mu_{\rm p}})\overline{P}_{\rm s,pd},\\
  \mu_{\rm s} &=& \alpha_{\rm s}(1-\frac{\lambda_{\rm p}}{\mu_{\rm p}})\overline{P}_{\rm s,sd}.
 \normalfont \end{eqnarray}

The stable-throughput region of the system can be obtained by formulating a constrained optimization problem similar to that discussed above for system $\mathcal{S}$. The optimization problem is expressed as:
\begin{eqnarray} \small\label{19000990}
    \underset{ f_{\rm s},\alpha_{\rm s}}{\max.} & &\alpha_{\rm s}(1-\frac{\lambda_{\rm p}}{\mu_{\rm p}})\overline{P}_{\rm s,sd} \notag\\ {\rm s.t.} & &     \lambda_{\rm p} \le \mu_{\rm p} \\& & \lambda_{\rm ps}\le\mu_{\rm ps} \notag\\& &0\le f_{\rm s}\le 1, 0\le \alpha_{\rm s}\le 1.
 \normalfont \end{eqnarray}

The relaying queue constraint can be rewritten as
\begin{eqnarray} \small
\lambda_{\rm p}\le  \mu_{\rm p} \frac{\overline{P}_{\rm s,pd}}{f _{\rm s} P_{\rm p,pd}\overline{P}_{\rm p,s} +\alpha_{\rm sp} \overline{P}_{\rm s,pd}}.
\label{ghj2}
 \normalfont \end{eqnarray}
 If the term $\frac{\overline{P}_{\rm s,pd}}{f _{\rm s} P_{\rm p,pd}\overline{P}_{\rm p,s} +\alpha_{\rm sp}\overline{P}_{\rm s,pd}}$ is less than the unity, the relaying queue stability constraint subsumes the primary queue stability constraint. If $\frac{\overline{P}_{\rm s,pd}}{f _{\rm s} P_{\rm p,pd}\overline{P}_{\rm p,s} +\alpha_{\rm sp}\overline{P}_{\rm s,pd}}$ is greater than the unity, then the primary queue stability constraint subsumes that of the relaying queue stability. Combining both cases, the constraint on $\lambda_{\rm p}$ which guarantees the stability of both the primary and relaying queues is
 \begin{eqnarray} \small
\lambda_{\rm p}\le  \mu_{\rm p} \min\Big\{\frac{\overline{P}_{\rm s,pd}}{f _{\rm s} P_{\rm p,pd}\overline{P}_{\rm p,s} +\alpha_{\rm sp} \overline{P}_{\rm s,pd}},1\Big\}.
\label{ghj3}
 \normalfont \end{eqnarray}

 For a given pair of $f_{\rm s}$ and $\alpha_{\rm s}$, the stability region of the network is given by Eqn. (\ref{boor}) at the top of the following page.
 \begin{table*}
\begin{equation}
\small \begin{split}
\label{boor}
   R(\hat{\mathcal{S}}^{(\rm NP)})&=\biggr\{ (\lambda_{\rm p},\lambda_{\rm s}): \lambda_{\rm s}< \alpha_{\rm s}\biggr(1-\frac{\lambda_{\rm p}}{\overline{P}_{\rm p,pd}+f_{\rm s} P_{\rm p,pd}\overline{P}_{\rm p,s}}\biggr)\overline{P}_{\rm s,sd} \biggr\}, \\ \text{with}\ \  \lambda_{\rm p}&\le ~(~\overline{P}_{\rm p,pd}~+~f_{\rm s} P_{\rm p,pd}\overline{P}_{\rm p,s}~)~\min~\Big\{~\frac{\overline{P}_{\rm s,pd}}{f _{\rm s} P_{\rm p,pd}\overline{P}_{\rm p,s}~+~\alpha_{\rm sp} \overline{P}_{\rm s,pd}},1~\Big\}.
     \\ \\
   \hline
   \normalsize \end{split}
\end{equation}
\end{table*}

Arranging the optimization problem (\ref{19000990}), the optimization problem becomes:
\begin{eqnarray} \small\label{190000990rr}
      \underset{ f_{\rm s},\alpha_{\rm s}}{\max.} & &\frac{\alpha_{\rm s}(\overline{P}_{\rm p,pd}+f_{\rm s} K-{\lambda_{\rm p}})}{\overline{P}_{\rm p,pd}+f_{\rm s} K},\notag \\ {\rm s.t.} & &  -f_{\rm s} K\le \overline{P}_{\rm p,pd}-\lambda_{\rm p} \notag \\& &0\le f_{\rm s}\le 1, 0\le \alpha_{\rm s}\le 1 \notag \\& &   \frac{f_{\rm s}}{\overline{P}_{\rm p,pd}+f_{\rm s} K-\lambda_{\rm p}}+\alpha_{\rm s}\frac{\overline{P}_{\rm s,pd} }{K\lambda_{\rm p}}\le\frac{\overline{P}_{\rm s,pd} }{K\lambda_{\rm p}},
 \normalfont \end{eqnarray}
where $K={P_{\rm p,pd}\overline{P}_{\rm p,s}}$. The optimization problem (\ref{190000990rr}) is non-convex. It can be solved via
a one-dimensional grid search over the optimal value of $f_{\rm s}$. Fixing $f_{\rm s}$, we have the following optimization problem:
\begin{eqnarray} \small\label{190000990}
      \underset{ \alpha_{\rm s}}{\max.} & &\alpha_{\rm s} \notag \\ {\rm s.t.} & &  f_{\rm s} \ge \frac{\lambda_{\rm p}-\overline{P}_{\rm p,pd}}{K}\notag \\& &0\le f_{\rm s}\le 1, 0\le \alpha_{\rm s}\le 1 \notag \\& &   \alpha_{\rm s}\le1-\frac{f_{\rm s}}{\overline{P}_{\rm p,pd}+f_{\rm s} K-\lambda_{\rm p}}\frac{K\lambda_{\rm p}  }{\overline{P}_{\rm s,pd}}.
 \normalfont \end{eqnarray}
The optimization problem is \textbf{linear} and can be readily solved. Note that we solve a family of convex optimization problems parameterized by $f_{\rm s}$. The optimal value of $f_{\rm s}$ is taken as the solution that yields the highest value of the objective function of the optimization problem (\ref{190000990rr}). The optimal value of $\alpha_{\rm s}$ for a fixed $f_{\rm s}$ is given by
\begin{equation}\label{222224}
    \alpha^*_{\rm s}=1-\frac{f_{\rm s}}{\overline{P}_{\rm p,pd}+f_{\rm s} K-\lambda_{\rm p}} \frac{K\lambda_{\rm p}  }{\overline{P}_{\rm s,pd}},
\end{equation}
    \begin{equation}\label{22222}
    \alpha^*_{\rm sp}=\frac{f_{\rm s}}{\overline{P}_{\rm p,pd}+f_{\rm s} K-\lambda_{\rm p}} \frac{K\lambda_{\rm p}  }{\overline{P}_{\rm s,pd}}.
\end{equation}
\noindent  It can be noted that the bound is proportionally increasing with $\alpha_{\rm s}$. That is, as the access probability increases, the stability bound expands. On the other hand, if the acceptance factor $f_{\rm s}$ increases, the probability of the primary queue being empty increases; hence, the bound expands. However, based on the relationship between $f_{\rm s}$ and $\alpha_{\rm s}$ as shown in Eqn. (\ref{222224}), $\alpha_{\rm s}$ is monotonic decreasing with $f_{\rm s}$ (see Appendix B for proof). Recall that $1-\alpha_{\rm s}=\alpha_{\rm sp}$ controls the service process of the relaying queue and $f_{\rm s}$ controls the arrival process to the relaying queue; hence, increasing $\alpha_{\rm s}$, which is equivalent to reducing the possibility of selecting the relaying queue for transmission, requires a reduction in $f_{\rm s}$ to maintain the relaying queue stability. Thus, there is a tradeoff between helping the PT and servicing the ST packets.

Substituting with $\alpha_s^*$ into the original optimization problem (\ref{190000990rr}), we get the following optimization problem:

\begin{eqnarray} \small\label{190000990rre}
      \underset{ f_{\rm s}}{\min.} & & \frac{f_{\rm s} }{\overline{P}_{\rm p,pd}-\overline{P}_{\rm s,pd}}
    \notag \\ {\rm s.t.} & &  -f_{\rm s} K\le \overline{P}l_{\rm p,pd}-\lambda_{\rm p} \notag \\& &0\le f_{\rm s}\le 1.
 \normalfont \end{eqnarray}
This optimization problem is exactly the one for the prioritized relaying system [optimization problem (\ref{1900099})]; thus, we conclude that random selection of queues does not expand the stability region of the system. In addition, this means that the prioritized relaying is the optimal strategy for the system where the ST is the only cooperative terminal in terms of stable-throughput region. However, we conjuncture that in terms of queueing delays of secondary packets, the random selection, $\hat{\mathcal{S}}^{(\rm NP)}$, would be better for the secondary queueing delay. This is because, in contrast to $\hat{\mathcal{S}}^{(\rm P)}$, the secondary packets under $\hat{\mathcal{S}}^{(\rm NP)}$ do not have to wait for the relaying queue to be emptied before getting services.

The optimal parameters for this system are given by

         \begin{itemize}
\item If $\overline{P}_{\rm p,pd}<\overline{P}_{\rm s,pd}$,
   \[ f^*_{\rm s} = \left\{ \begin{array}{ll}
   [0,1]& \mbox{if $\lambda_{\rm p} =0$} ;\\
         1 & \mbox{if $0<\lambda_{\rm p} \le \overline{P}_{\rm s,pd}$}.\end{array} \right. \]

             $$\alpha^*_{\rm s}=1-\frac{1}{\overline{P}_{\rm p,pd}+ K-\lambda_{\rm p}} \frac{K\lambda_{\rm p}  }{\overline{P}_{\rm s,pd}}$$

  $$  \alpha^*_{\rm sp}=\frac{1}{\overline{P}_{\rm p,pd}+ K-\lambda_{\rm p}} \frac{K\lambda_{\rm p}  }{\overline{P}_{\rm s,pd}}.$$

        \item If $\overline{P}_{\rm p,pd}>\overline{P}_{\rm s,pd}$,
   \[ f^*_{\rm s} = \left\{ \begin{array}{ll}
   [0,1]& \mbox{if $\lambda_{\rm p} =0$} ;\\
         0 & \mbox{if $0<\lambda_{\rm p} \le \overline{P}_{\rm p,pd}$}.\end{array} \right. \]

    $$\alpha^*_{\rm s}=1, \ \alpha^*_{\rm sp}=0.$$

               \item If $\overline{P}_{\rm p,pd}=\overline{P}_{\rm s,pd}$, the problem is reduced to a feasibility problem. However, we will obtain set of fractions $f_{\rm s}$ that satisfies the domain of the objective function and the constraints. Thus, the optimal set of $f_{\rm s}$ is given by

         \begin{equation}
\small \begin{split}
 f_{\rm s}^*=[0,1].
 \normalsize \end{split}
\end{equation}
The optimal access probabilities
        \begin{equation}
\small \begin{split}
 \alpha^*_{\rm s}&=1-\frac{f_{\rm s}}{\overline{P}_{\rm p,pd}+f^*_{\rm s} K-\lambda_{\rm p}} \frac{K\lambda_{\rm p}  }{\overline{P}_{\rm s,pd}}, \ \\ \alpha^*_{\rm sp}&=\frac{f_{\rm s}}{\overline{P}_{\rm p,pd}+f^*_{\rm s} K-\lambda_{\rm p}} \frac{K\lambda_{\rm p}  }{\overline{P}_{\rm s,pd}}.
  \normalsize \end{split}
\end{equation}
One of the solutions is $f_{\rm s}=0$. In such case, the optimal values of $\alpha_{\rm s}$ and $\alpha_{\rm ps}$ are $\alpha_{\rm s}^*=1$ and $\alpha_{\rm ps}^*=0$, respectively.

         \end{itemize}

The stability of $\hat{\mathcal{S}}^{(\rm NP)}$ is exactly that of $\hat{\mathcal{S}}^{(\rm P)}$. If $\overline{P}_{\rm p,pd}<\overline{P}_{\rm s,pd}$, the stability region is given by
\begin{equation}
\small \begin{split}
   R(\hat{\mathcal{S}}^{(\rm NP)})&=   R(\hat{\mathcal{S}}^{(\rm P)}).
%
 \normalsize \end{split}
\end{equation}
\section{Numerical Results}\label{numerical}
In this section, we provide the solution of the optimization problems considered in this paper.
\begin{figure}[t]
  \centering
  \includegraphics[width=1 \columnwidth]{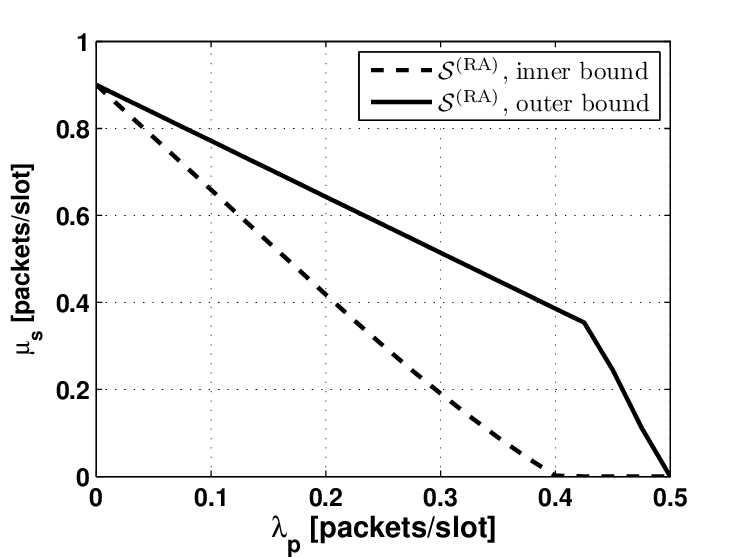}\\
  \caption{Inner and outer bounds for $\mathcal{S}^{\left(\rm RA\right)}$.}\label{fig1}
\end{figure}
The inner (the union of the dominant systems stability regions) and the outer (intersection of the proposed outer bounds) bounds of $\mathcal{S}^{\left(\rm RA\right)}$ are depicted in Fig. \ref{fig1}. The parameters used to generate the figure are: $P_{\rm p,pd}=1$
$P_{\rm s,sd}=0.1$, $P_{\rm p,s}=0.3$, $P_{\rm s,pd}=0.2$, $P_{\rm sd,pd}=0.2$, $P_{\rm p,sd}=0.3$, $P^{\rm sd}_{\rm s,pd}=0.68$, and $P^{\rm s}_{\rm sd,pd}=0.68$. As is obvious, the outer bound contains the inner bound. The relationship between $\mu_{\rm s}$ and $\lambda_{\rm p}$ is monotonic.

\begin{figure}
  \centering
  \includegraphics[width=1 \columnwidth]{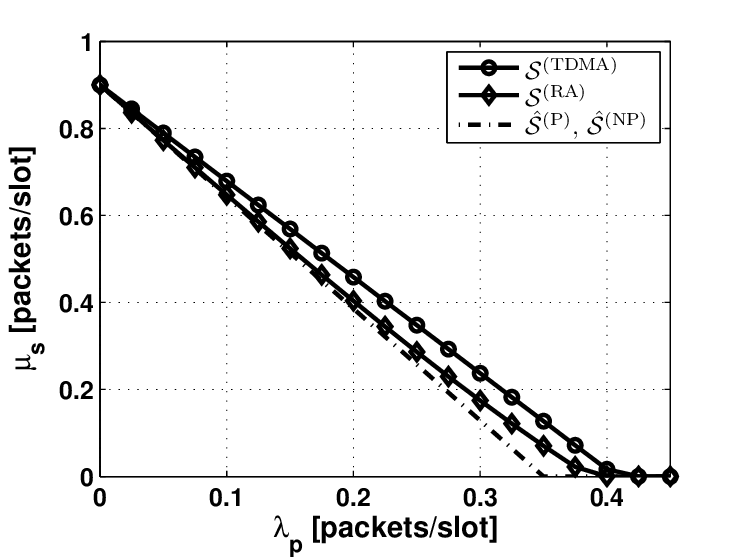}\\
  \caption{The stable throughput region of $\mathcal{S}^{\left(\rm TDMA\right)}$, $\mathcal{S}^{\left(\rm RA\right)}$, $\hat{\mathcal{S}}^{\left(\rm P\right)}$ and $\hat{\mathcal{S}}^{\left(\rm NP\right)}$.}\label{fig2}
\end{figure}

\begin{figure}
  \centering
  \includegraphics[width=1 \columnwidth]{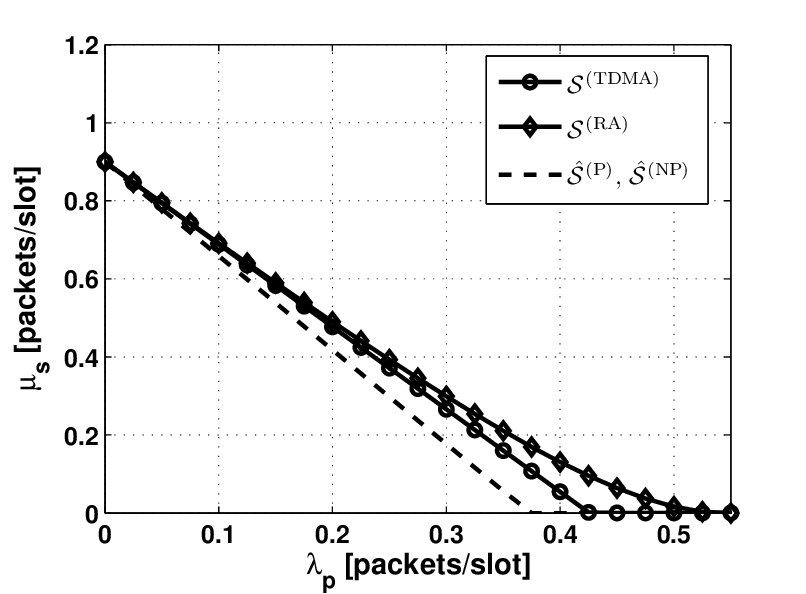}\\
  \caption{The stable throughput region of $\mathcal{S}^{\left(\rm TDMA\right)}$, $\mathcal{S}^{\left(\rm RA\right)}$, $\hat{\mathcal{S}}^{\left(\rm P\right)}$ and $\hat{\mathcal{S}}^{\left(\rm NP\right)}$ for the case of strong MPR at the PR.}\label{strongmpr}
\end{figure}

\begin{figure}
  \centering
  \includegraphics[width=1\columnwidth]{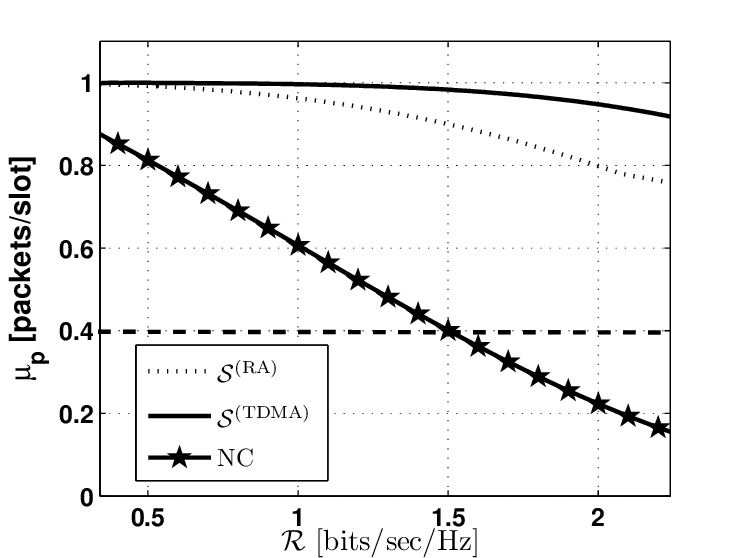}\\
  \caption{Maximum primary mean service rate of $\mathcal{S}^{\left(\rm TDMA\right)}$ and $\mathcal{S}^{\left(\rm RA\right)}$ versus $\mathcal{R}$. The non-cooperation (NC) case is also plotted for comparison.}\label{fig3}
\end{figure}

The stability regions of $\mathcal{S}^{\left(\rm TDMA\right)}$, $\mathcal{S}^{\left(\rm RA\right)}$, $\hat{\mathcal{S}}^{\left(\rm P\right)}$ and $\hat{\mathcal{S}}^{\left(\rm NP\right)}$ are depicted in Figs. \ref{fig2} and \ref{strongmpr}. The parameters used to generate Fig. \ref{fig2} are the same as Fig. \ref{fig1}. For Fig. \ref{strongmpr}, we use the exact same parameters of Fig. \ref{fig2} with $P^{\rm sd}_{\rm s,pd}=P_{\rm s,pd}=0.2$ and $P^{\rm s}_{\rm sd,pd}=P_{\rm sd,pd}=0.2$. These values are chosen to show impact of the strong MPR capability of the PR on the stability region. Since $P_{\rm p,pd}=1$, the mean service rate of the primary queue without cooperation is zero. This means that the primary packets will not get service. Hence, the stability region of the primary-secondary network without cooperation is the $(\lambda_{\rm p}=0,\lambda_{\rm s}=0)$. The proposed protocols provide significantly higher stability regions than the non-cooperation case. From Fig. \ref{fig3}, it is noted that $\mathcal{S}^{\left(\rm TDMA\right)}$ system has the highest stable-throughput region among the considered systems. The RA system $\mathcal{S}^{\left(\rm RA\right)}$ has the second best performance among the considered systems. In Fig. \ref{strongmpr}, we show the case of strong MPR capability at the PR. From the figure, the RA system $\mathcal{S}^{\left(\rm RA\right)}$ outperforms the TDMA system $\mathcal{S}^{\left(\rm TDMA\right)}$ and the special case systems $\hat{\mathcal{S}}^{\left(\rm P\right)}$ and $\hat{\mathcal{S}}^{\left(\rm NP\right)}$. As proved in our analysis, Figs. \ref{fig2} and \ref{strongmpr} show that the stability regions of $\hat{\mathcal{S}}^{\left(\rm P\right)}$ and $\hat{\mathcal{S}}^{\left(\rm NP\right)}$ are coincide. Furthermore, the stability regions of $\hat{\mathcal{S}}^{\left(\rm P\right)}$ and $\hat{\mathcal{S}}^{\left(\rm NP\right)}$ are affine; hence, convex. This follows our analysis.

The maximum primary mean service rate versus $\mathcal{R}=\mathcal{R}_{\rm p}=b/W/T$ is plotted in Fig. \ref{fig3}. The maximum $\mu_{\rm p}$ for $\mathcal{S}^{\left(\rm TDMA\right)}$ is higher than $\mathcal{S}^{\left(\rm RA\right)}$. Moreover, both schemes provide higher primary service rates than the non-cooperation (NC) case. Let $\gamma_{\rm j,k} =\mathbb{P}_{\rm j}/\mathcal{N}_{\rm k}$ be the received signal-to-noise ratio (SNR) at receiver ${\rm k}$ when the channel gain $h_{\rm j,k}$ is unity. The parameters used to generate the figure are: $\gamma_{\rm sd,pd} \sigma^2_{\rm sd,pd}=10$, $\gamma_{\rm s,pd} \sigma^2_{\rm s,pd}=10$,
$\gamma_{\rm p,pd} \sigma^2_{\rm p,pd}=2$, $\gamma_{\rm p,s} \sigma^2_{\rm p,s}=10$, $\gamma_{\rm p,sd} \sigma^2_{\rm p,sd}=10$, $\gamma_{\rm s,sd} \sigma^2_{\rm s,sd}=8$, $\tau=0.1 T$ and $\lambda_{\rm p}=0.4$ packets/slot.

\begin{figure}
  \centering
  \includegraphics[width=1 \columnwidth]{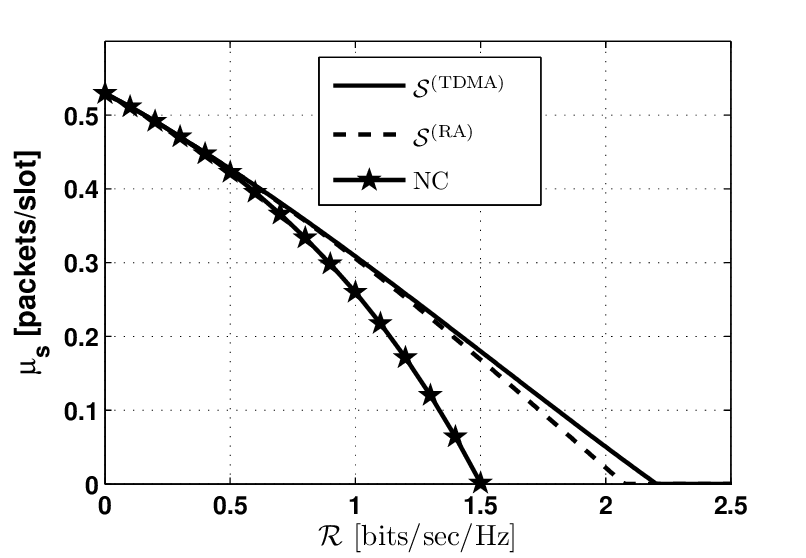}\\
  \caption{Maximum secondary stable throughput of $\mathcal{S}^{\left(\rm TDMA\right)}$ and $\mathcal{S}^{\left(\rm RA\right)}$ versus $\mathcal{R}$. The non-cooperation (NC) case is also plotted for comparison.}\label{fig4}
\end{figure}

Fig. \ref{fig4} shows the maximum secondary stable throughput versus $\mathcal{R}$. The parameters used to generate the figure are exactly those of Fig. \ref{fig3}. As shown in the figure, the maximum secondary throughput decreases with $\mathcal{R}$. This is because as $\mathcal{R}$ increases, the outage probabilities of all links increase as well. Hence, the probabilities of correct packets decoding and the service rates decrease. This fact follows Eqns. (\ref{145}) and (\ref{193}).

From the figures, we conclude the following.
 \begin{itemize}
 \item The envelopes of the stability regions are monotonically decreasing with the mean arrival rate of the PT, $\lambda_{\rm p}$. This is because as the the primary arrival rate increases, the probability of the primary queue being empty vanishes. Hence, the probability of having an empty slot for second access tends to zero.
     \item The TDMA-based system outperforms all systems at low MPR capability at the PR. On the other hand, for high MPR, the RA-based system outperforms all systems.
         \item The queues rates decrease with $\mathcal{R}$. This is because the outage probabilities increase; hence, the service rates decrease.
          \item It is noted that the feasible range of the primary arrival rate expands due to cooperation. This is because the primary packets can be served due to existence of either one or two relay station(s) in case of $\hat{\mathcal{S}}^{\left(\rm P\right)}$ and $\hat{\mathcal{S}}^{\left(\rm NP\right)}$ or $\mathcal{S}^{\left(\rm RA\right)}$ and $\mathcal{S}^{\left(\rm TDMA\right)}$, respectively. The relaying stations help in delivering the primary packets during the silence periods of the PT; hence, increase the probability of servicing the arrived packets to the primary queue without wasting either frequency bandwidth or time slots.
              \end{itemize}

\section{Conclusion}\label{conc}
In this paper, we have addressed the impact of cooperation between the primary and secondary systems on their performance from the network-layer standpoint. We have investigated the maximum stable-throughput of $\mathcal{S}$, $\hat{\mathcal{S}}^{(\rm P)}$, and $\hat{\mathcal{S}}^{(\rm NP)}$ systems. In $\mathcal{S}$, the cognitive system with its transmitter-receiver pair senses the channel for idle channel resources and exploits them to relay the undelivered packets of the primary user or to serve its own traffic. To manage the multiple access nature of the channel, we have proposed two multiple access schemes; namely RA and TDMA schemes. The TDMA-based system $\mathcal{S}^{(\rm TDMA)}$ has shown to provide the highest stable-throughput for the secondary system in case of low MPR capability. On the other hand, the RA-based transmission system $\mathcal{S}^{(\rm RA)}$ has shown to be the best system in case of strong MPR capability. This means that at high MPR capability, the RA scheme, which requires less cooperation between the ST and the SR than the TDMA scheme to be implemented, provides higher throughput region. We have proposed two special cases of $\mathcal{S}$. In the first special case, denoted by $\hat{\mathcal{S}}^{(\rm P)}$, the ST assigned higher priority to the relaying queue, if the relaying queue is empty, it transmits a packet from its own queue. In the second special case, denoted by $\hat{\mathcal{S}}^{(\rm NP)}$, the ST uses the silence periods of the primary user to serve a packet from one of its queues. The selection of queues is random with certain adjustable probability. The analysis has shown that both $\hat{\mathcal{S}}^{(\rm NP)}$ and $\hat{\mathcal{S}}^{(\rm P)}$ provide the exact same stability regions. This means that when only the ST cooperates with the PT, then the optimal selection policy between queues will be to assign the highest priority of transmission to the relaying packets. The results have revealed the significant performance of $\mathcal{S}$ over $\hat{\mathcal{S}}^{(\rm NP)}$ and $\hat{\mathcal{S}}^{(\rm P)}$. In other words, the maximum stable-throughput regions of $\hat{\mathcal{S}}^{(\rm P)}$ and $\hat{\mathcal{S}}^{(\rm NP)}$ are always subsets of the maximum stable-throughput of $\mathcal{S}$.
\section*{Appendix A}
Under Rayleigh fading channels, the outage probability between terminal ${\rm j}$ and terminal ${\rm k}$ when there is a transmission caused by node $\ell$ \cite{erph} is given by
 \begin{equation}\label{193x}
 \small \begin{split}
 P^\ell_{j,k}={\rm Pr}\{O_{j,k}|\mathcal{T}_{\ell}\}&={\rm Pr}\{\frac{\mathbb{P}_j |h_{j,k}|^2}{\mathbb{P}_\ell |h_{\ell,k}|^2+\mathcal{N}_{\rm k}}\le \gamma^{\left(\rm th\right)}_{\rm j}\}\\&\!=\!1\!-\!\frac{1}{1\!+\!\frac{\mathbb{P}_\ell\gamma^{\left(\rm th\right)}_{\rm j} }{\mathbb{P}_j }\frac{\sigma^2_{\ell,k}}{\sigma_{j,k}^2}}\exp\bigg(\!-\!{\frac{\gamma^{\left(\rm th\right)}_{\rm j} \mathcal{N}_{\rm k}}{\mathbb{P}_j \sigma_{j,k}^2}}\bigg)
 \normalsize \end{split}
\end{equation}
where $\{\mathcal{T}_{\ell}\}$ is the event that the terminal $\ell$ is transmitting a packet. After some mathematical manipulations, it can shown that the probability of correct packet reception in case of interference is given by
 \begin{eqnarray*}\label{193xx}
 \overline{P^\ell_{j,k}}=\!\frac{\overline{P}_{j,k}}{1+\frac{\mathbb{P}_\ell\gamma^{\left(\rm th\right)}_{\rm j} }{\mathbb{P}_j }\frac{\sigma^2_{\ell,k}}{\sigma_{j,k}^2}}.
\end{eqnarray*}
Note that for strong MPR, $\overline{P^\ell_{j,k}}\approx  \overline{P}_{j,k}$. This happens when $\frac{1}{1+\frac{\mathbb{P}_\ell\gamma^{\left(\rm th\right)}_{\rm j} }{\mathbb{P}_j }\frac{\sigma^2_{\ell,k}}{\sigma_{j,k}^2}}\approx 1$; hence, $\gamma^{\left(\rm th\right)}_{\rm j}\approx 0$.
\section*{Appendix B}
In this Appendix, we prove that $\alpha_{\rm s}$ is a monotonically nonincreasing function of $f_{\rm s}$. From (\ref{222224}), we have
\begin{equation}\label{222200002}
    \alpha^*_{\rm s}=1-\frac{f_{\rm s}}{\overline{P}_{\rm p,pd}+f_{\rm s} K-\lambda_{\rm p}} \frac{K\lambda_{\rm p}  }{\overline{P}_{\rm s,pd}}
\end{equation}
The first derivative is given by
\begin{equation}\label{222200002}
    d\frac{\alpha^*_{\rm s}}{df_{\rm s}}=-\frac{\overline{P}_{\rm p,pd}}{\bigg(\overline{P}_{\rm p,pd}+f_{\rm s} K-\lambda_{\rm p}\bigg)^2} \frac{K\lambda_{\rm p}  }{\overline{P}_{\rm s,pd}}
\end{equation}
As $f_{\rm s}$ varies over its domain $[0,1]$, the first derivative is strictly negative. This implies that $\alpha_{\rm s}^*$ is a monotonically nonincreasing function of $f_{\rm s}$.
\section*{Acknowledgement}
This research work is funded by Qatar National Research Fund (QNRF) under grant number NPRP 09-1168-2-455.
\bibliographystyle{IEEEtran}
\bibliography{IEEEabrv,term_bib}
\balance
\end{document}